\documentclass[twocolumn,aps,prl,showpacs,longbibliography]{revtex4-1}
\usepackage[T1]{fontenc}
\usepackage[utf8]{inputenc}
\usepackage{amsmath}
\usepackage{amssymb}
\usepackage{graphicx}
\usepackage{esint}
\usepackage[unicode=true,
 bookmarks=false,
 breaklinks=false,pdfborder={0 0 1},backref=section,colorlinks=false]
 {hyperref}
\hypersetup{
 colorlinks,citecolor=blue,linkcolor=blue,urlcolor=blue,hyperfootnotes=true}
\begin{document}
\title{{Charge order and emergent symmetries in cuprate
superconductors}}
\author{{C. Pépin}}
\affiliation{{Institut de Physique Théorique, Université Paris-Saclay,
CEA, CNRS, F-91191 Gif-sur-Yvette, France}}
\author{{H. Freire}}
\affiliation{{Instituto de Física, Universidade Federal de Goiás,
74.001-970, Goiânia-GO, Brazil}}
\begin{abstract}
{In this paper, we present our studies of the phase
diagram of the cuprate superconductors performed in recent years.
We describe how a few field-theoretical concepts can be used to account
for the puzzling properties of these compounds. Starting with a short
exposition of recent experimental developments, we then introduce
the concept of an emergent SU(2) symmetry, which rotates between the
superconducting and charge order parameters. We describe a solvable
model for which this symmetry turns out to be exact and derive the
corresponding effective field theory in terms of a non-linear $\sigma$
model. We then turn to the experimental consequences of our model
with the presence of skyrmions inside the vortex core and an elegant
account of the $B$-$T$ phase diagram. Next, we move on to the second
concept, which is the one of an emergent U(1) gauge field in the theory.
Within this more general framework, the pseudogap transition can be
viewed as a confining/deconfining transition. Experimental consequences
for a few spectroscopic probes like inelastic neutron scattering,
Raman spectroscopy, phonon line in x-ray scattering and angle-resolved
photoemission experiments are studied. This theory suggests the presence
of incoherent bosons at moderate temperatures in the phase diagram
of the cuprates, which would undergo a jamming transition at optimal
doping \cite{Pierfrancesco2022}. Consequences of the presence of
such bosons on the transport properties of these materials are examined. }
\end{abstract}
\maketitle

\section{Introduction\label{subsec:Introduction}}

\begin{figure}
{\includegraphics[width=8cm]{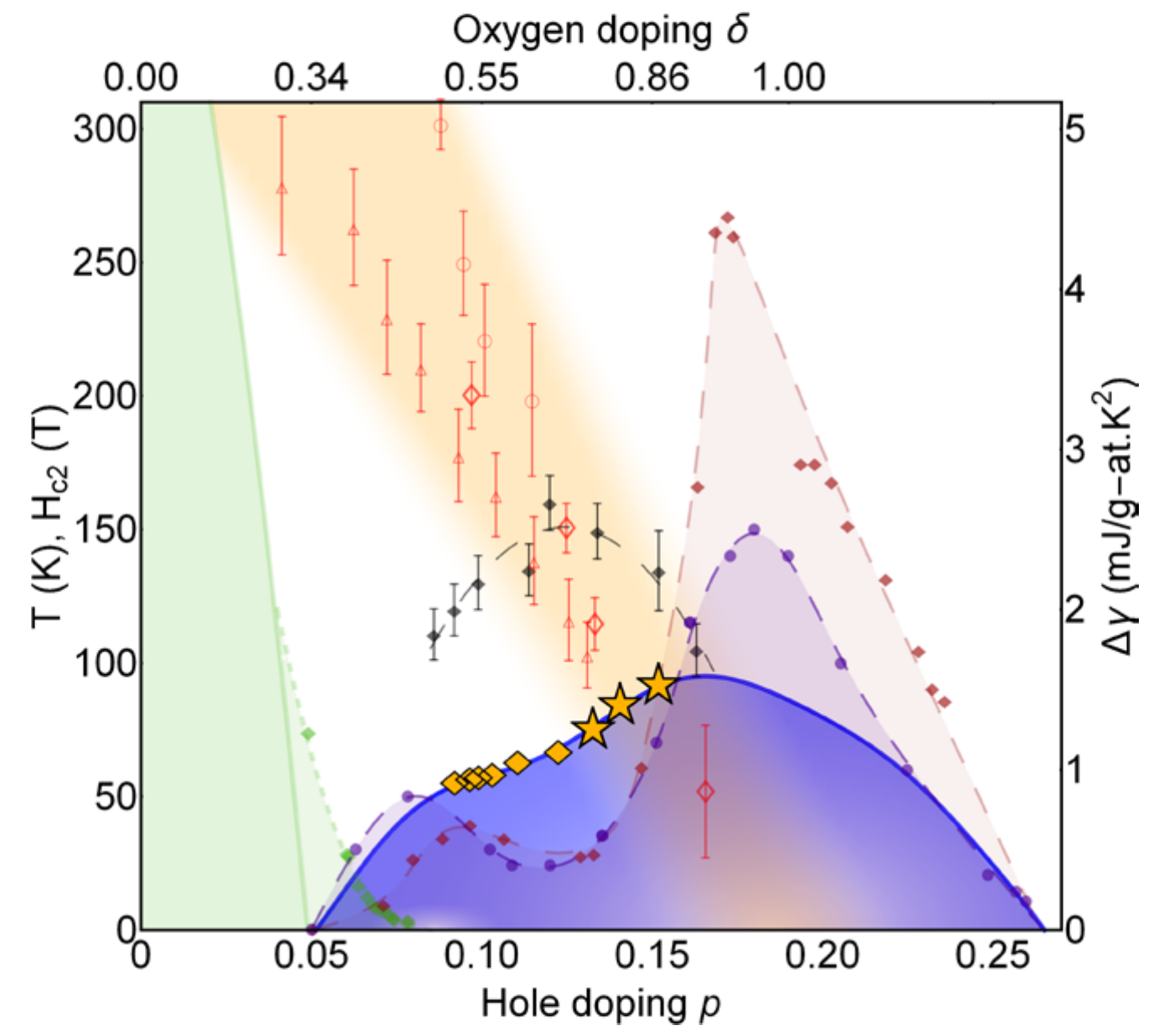} \caption{\label{Fig1}Experimental phase diagram of the cuprate superconductors from Ref.~\cite{Ramshow15}.
We note the Mott insulator {on} the {left-hand} side, the Fermi
liquid region {on} the {right-hand} side and {the PG} phase
{is in the underdoped part of the phase} diagram.}
}
\end{figure}

{The field of cuprate superconductivity has been
considered for more than forty years now as the Rosetta stone for
condensed matter physics~\cite{Lee06,Norman03,LeHur:2009iw,Anderson87,Fradkin15,Tsvelik:2007cr,SenthilLee09,Rice12}.
It is the subject in which the highest number of papers have been
published in the research area of strongly correlated systems. The
reason for this intense activity might elude a physicist outside the
field. When giving a first look at the phase diagram of the cuprates
as depicted in Fig. \ref{Fig1}, however, we start to see very unusual
features of this material. This family of copper oxides has the highest
superconducting (SC) temperature $T_c$ (without external pressure applied)
and, although the phenomenon of superconductivity expels magnetic
flux, the SC phase is situated in the vicinity of an antiferromagnetic
(AF) phase with a transition temperature three times higher than the
SC one. More surprisingly, it is also situated in the vicinity of
a Mott insulating phase.}

{However, these are not the only mysteries hidden
in this phase diagram. A pseudogap (PG) phase also exists below a
temperature $T^{*}$, where a gap is seen to open at the Fermi surface
in the antinodal (AN) region of the Brillouin zone, leaving Fermi
arcs in the nodal region (see, e.g., \cite{Norman03}). This phase,
which was first evidenced in the underdoped region of the hole-doped
compounds by nuclear magnetic resonance (NMR) experiments ~\cite{Alloul89,Warren89},
remains a mystery eluding understanding for forty years. On the right-hand
side of the phase diagram, a quasi-metallic phase exists which obeys
the standard laws of the Fermi liquid theory for electric transport
in metals~\cite{NozieresPines}. But again around optimal doping,
in between the PG and Fermi liquid phases, we find what resembles
one of the most strongly entangled fixed point of condensed matter
theory~\cite{zaanen2019planckian}. Here, experiments show very unusual
electric and thermal transport properties, which defy the standard
theories of transport valid within the Fermi liquid paradigm (see,
e.g., \cite{hussey2008phenomenology,hussey2013generic,husseycupratescriticality}).
Indeed, this phase denoted by ``strange metal phase'' shows $T$-linear
resistivity up to a thousand of Kelvins, whereas the Wiedemann-Franz
law is satisfied (see, e.g., \cite{Michon_Taillefer} and references
therein).}

{Recently, the field has been revived by the experimental
discovery of charge order (CO) in the underdoped regime of cuprates.
It started with the observation by scanning tunneling microscopy (STM)
that modulations exist inside vortices when a magnetic field is applied~\cite{Hoffman02}.
These modulations were soon enough seen by NMR experiment~\cite{Wu11,Wu13a,Wu14},
which also was able to confirm that the signal is non-magnetic, suggesting
CO as the main source. Meanwhile, a very important quantum oscillation
(QO) experiment showed that the Fermi surface is reconstructed in
the underdoped region of the phase diagram, and that small electron
pockets are present~\cite{Doiron-Leyraud07,Sebastian12,Doiron-Leyraud15}.
The origin of the reconstruction was soon linked to the presence of
CO, which enables the creation of small electron pockets detectable
by QO. X-ray experiments and resonant x-rays on the copper site confirmed
the presence of CO in this system \cite{Blackburn13a,Chang12,Chang16,LeTacon11,Comin14,Comin:2015ca,Comin:2015vc,Comin15,Ghiringhelli12,Blanco-Canosa13,Blanco-Canosa14,Chaix:2017fs}
and also explored, concomitantly with ultrasound experiments \cite{LeBoeuf13,Chang:2016gz},
the $B$-$T$ phase diagram when a magnetic field is applied (see
Fig. \ref{Fig6} in this paper). Here, we will also focus on a few
other experimental probes prominent in the study of the underdoped
region, like Raman scattering~\cite{Benhabib:2015ds,Benhabib:2015kw,Loret19},
angle-resolved photoemission ~\cite{Damascelli03,Vishik18,Vishik:2010fn,Vishik:2010tc,Vishik:2012cc},
neutron and muon scattering (elastic and inelastic)~\cite{Fauque06,Eschrig:2006ky,Mangin_Thro14,Mangin_Thro17,ManginThro:2015fg,ManginThro:2014js,Baledent11,Bourges11,Hayden04,Hinkov04,Hinkov07,Dai99},
ultrasound and x-ray observation of phonons ~\cite{Blackburn13b,LeTacon14},
and probes of the electric and thermal transport~\cite{AbdelJawad:2006df,AbdelJawad:2007el,Barisic:2015tg,Bozovic04,chen2019incoherent,CyrChoiniere:2018ed,Grissonnanche19,hartnoll2015theory,hartnoll2018holographic,Homes04,Homes05,hussey2008phenomenology,hussey2003coherent,Hwang94,lobo2011optical,RossMcKenzie12,RossMcKenziePRL11,Tallon01,Tallon95,Wang:2006fa,Zaanen14,RullierAlbenque:2000fl,RullierAlbenque:2007bm,RullierAlbenque:2011ji}.
We will describe broadly these experiments, but keeping focused on
the observation of charge order in the underdoped regime.}

{This very mysterious phase diagram (Fig. \ref{Fig1})
resisted theoretical approaches for a very long time, but also led
experimentalists and theoreticians to propose new ideas and unconventional
scenarios \cite{abanov03,agterberg2020physics,Altman02,Anderson87,Atkinson15,Atkinson:2015kk,Barisic:2015tg,Benfatto00,Benfatto:2007df,Berg:2008ii,Berg09,Caprara:2016gs,Caprara:2016vh,Carbotte:2011ip,Carlson:2004hn,Chakraborty19,Chakravarty01,Chatterjee19,Chowdhury14,Civelli05,Coleman96,Corboz14,Demler04,Demler95,Deutscher:1999jy,Efetov13,Eremin:2005ba,ErezBerg20,Ferraz:2015voa,Freire:2015kg,Ghosal98,hartnoll2015theory,Honerkamp:2004dx,Ioffe98,Kivelson:2002ug,Konik06,Kotliar88a,Kotliar88b,Lee06,Lee14,LeHur:2009iw,Norman03,Norman:2007eq,Onufrieva:2017jf,Parmekanti01,Pepin98,Rice12,RossMcKenzie12,Sachdev19,Senthil:2000eb,SenthilLee09,Stojkovic97,VojtaRosch08,Zaanen89,zaanen2019planckian}.
We would like to highlight here three main types of ideas that have
been influential over the years for the study of the cuprate phase
diagram. First of all, when we look at the phase diagram of the cuprates,
we see that we are close to a Mott transition, which occurs at zero
doping in Fig. \ref{Fig1}. This metal-insulator transition implies
a large Coulomb repulsion on each site, which is of the order of 1 eV.
At half filling, the electrons are totally localized on the different
sites of the lattice, forming an insulator. This formation of an insulator
is accompanied by an antiferromagnetic order that persists until a
temperature of about 700 K. The presence of strong Coulomb energy,
which is the largest energy scale of the system, induces strong correlations
on the electrons, to the point of giving the name of strongly correlated
electron system to this part of solid state physics. In this first
approach, the formation of the PG is due to strong Coulomb interactions,
which typically induce a ``fractionalization'' of the electron into
spinon and holon~\cite{Lee06,Kotliar88b}. In typical theories, the
process is described by an emergent gauge symmetry, typically U(1)
or SU(2), and the PG is attributed to spinon pairing above $T_{c}$.
This type of theories invoking a form of spin liquid or spinon pairing
in the PG phase of the cuprates was first introduced intuitively by
P. W. Anderson under the name of ``resonating valence bond'' state~\cite{Anderson87,Balents05,Chakravarty01,Coleman84,Kotliar88a,Kotliar88b,SenthilLee09,Stanescu06,Stanescu07,Tsvelik:2007cr,Ferraz:2015voa,Japaridze:2002ie,Parmekanti01},
which involves entangled pairs of spins.}

{The second striking aspect when looking at the phase
diagram of the cuprates is that, starting from large enough oxygen
doping, we find the phenomenology of the Fermi liquid, which describes
conventional metals at low temperature~\cite{NozieresPines}. This
has given rise to the idea that a quantum critical point (QCP), or
a zero-temperature phase transition, is hidden under the superconducting
dome and that the PG phase is related to a ``broken'' symmetry of
the system corresponding to this QCP~\cite{Ramshow15,Tallon01}.
This situation where a QCP is under the superconducting dome is quite
common in the physics of strongly correlated quantum materials~\cite{Abanov00,abanov03,Abanov04,Chubukov:2007gk,Monthoux:2007ha,VojtaRosch08,Zaanen89,Zaanen:1998cl,Wang:2002ke}.
To name just one example out of many, it can be found, e.g., in heavy fermion compounds (see, e.g.,
\cite{gegenwart2008quantum}).}

{The last angle of approach to describe the cuprates
is to consider that the intermediate oxygen doping phase is a phase
in which many fluctuations are present~\cite{EmeryVJ:1995dr}. There
are several physical reasons for this. On the one hand, the system
is very anisotropic and quasi-bidimensional, which induces strong
quantum fluctuations. On the other hand, the PG phase is close to
a metal-insulator transition (the Mott transition) and, during such
a localization transition, the corresponding phases of the different
particles fluctuate. This line of thought has been pursued, e.g.,
by considering the fluctuations of the phase of the superconducting
pairs~\cite{Grilli:1990uf,Kivelson:1998ir,Kivelson:2002ug,EmeryVJ:1995dr,EmeryVJ:1995es,EmeryVJ:1999tx,hartnoll2015theory,Nayak:2000hg,Zhang:2002hz,Zhang:1997ew,Zhang:1999fe,Zhang1990,Demler04,Demler95}.
While it gives a solid phenomenology, in particular, for the closure
of the PG observed by angle-resolved photoemission spectroscopy~\cite{Chatterjee06,Norman97,Norman:1995dd,Norman:1998hk,NormanKanigel07},
it is now consensual to think that the phase of the superconducting
pairs alone does not extend to the PG temperature $T^{*}$, but rather
to a few tens of degrees above $T_{c}$~\cite{Chang2010}.}

{}

{This paper gives an alternative view on the PG phase
and the strange metal phase of the cuprates through a set of original
ideas with the potential to explain a large body of experimental data.
It is based on the consideration that charge order in the underdoped
region is central to explain the formation of the PG phase. Most of
these ideas were originally discussed in collaborations with K. B.
Efetov, where the concepts of emergent symmetry rotating from the
charge order to the SC state were developed.}

{The paper starts with the concept of an emergent
SU(2) symmetry in Section \ref{sec:Emergent SU(2) symmetry}. Applications
to the observation of modulations in the vortex core, description
of the $B-T$ phase diagram, Raman spectroscopy and inelastic neutron
scattering (INS) experiments are given. We then turn to the concept
of fractionalization of a pair density wave (PDW) in Section \ref{sec:Fractionalized-Pair-Density}.
Applications to angle-resolved photoemission experiment (ARPES), phase
locking of the charge density wave (CDW), phonon softening below $T_{c}$
and the description of the possibility of loop currents are given
in this section. In Section \ref{sec:Strange-Metal}, we turn to a
description of the strange metal phase within this scenario based
on the possibility for charged bosons to be present in a large part
of the phase diagram. We end with a discussion of our results and
a comparison of our scenario with other attempts to account for the
effect of the charge order in the underdoped phase of the cuprates
in Section \ref{sec:Discussion}. We conceived this paper as a small
tribute and review of the work done in collaboration with K. B. Efetov.
It does not have the vocation of full novelty since most of the ideas
presented here are already published, but the section about the strange
metal has new concepts and some new results. It does not claim to
be exhaustive either, nor to be the final solution of the problem,
but simply to offer an alternative scenario for the interpretation
of the phase diagram of those compounds, compared to the various ones
presented over the years.}

\section{{Emergent SU(2) symmetry}}

{\label{sec:Emergent SU(2) symmetry}}

\subsection{{Effective field theory}}

{In order to account for the experimental observations,
and the presence of CO inside the PG phase, we introduced the concept
of emergent SU(2) symmetry. This concept is not new: It amounts to
considering rotations from the SC pairing state to another state with
different symmetry. For example, one can rotate the SC state into
the AF state within the SO(5) symmetry group~\cite{Demler95,Demler04,Zhang:1997ew,Zhang:1999fe}.
This was an obvious choice at the time, since the AF phases and the
SC phases are the most visible physical states in the phase diagram
of these compounds. This led the authors of this theory to produce
an original explanation for the magnetic resonance seen in inelastic
neutron scattering (INS) around the phase diagram~\cite{Demler95}.
Indeed it has been shown experimentally that such a resonance is critical
inside the AF phase, and progressively gets massive when the hole
doping is increased~\cite{Demler95,Chubukov:479596}. A criticism
for this idea came from the observation that the resonance of the
SO(5) theory forms an antibonding state and thus is situated at an
energy too high to account for the experimental data~\cite{NormanChub01}.
The deep reason why this idea although brilliant seem to be failing
to account for the data is that in the AF phase and close to the SC
phase of the phase diagram there is of course a Mott insulating phase.
Namely, the two states of AF and SC are not close enough in energy
to allow for the rotation between these two phases to be really possible~\cite{NormanChub01}.}

\begin{figure}
{\includegraphics[width=8cm]{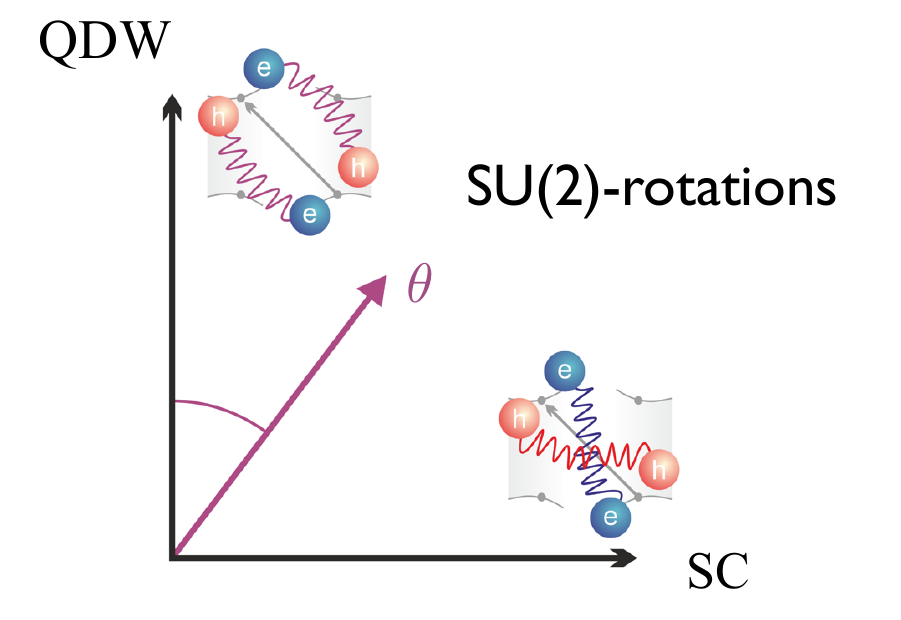} \caption{\label{Fig2}$\eta$-modes rotating from the SC state to the CDW state
and representing the SU(2) emergent symmetry{. From} ~\cite{Efetov13}.
{QDW in this figure stands for quadrupolar density
wave, which highlights the additional quadrupolar symmetry of the
CDW in the context of the model of Ref. \cite{Efetov13}}.}
}
\end{figure}

{Another symmetry known to be present in the phase
diagram of the cuprates is the SU(2) symmetry, which rotates between
the particle-particle wave function and particle-hole wave function~\cite{Kotliar88b,Lee98}.
This symmetry has been extensively studied close to half-filling where
it is exact, especially within strong coupling techniques that include
slave-bosons methods. Our idea was to build on this exact symmetry
and consider a rotation from the SC state to the observed CO~\cite{Zhang1990}.
In real space, the rotation involves two $\eta${-}operators schematically
represented in Fig. \ref{Fig2}, $\eta^{\dagger}=1/2\sum_{\sigma}\left(c_{i\sigma}^{\dagger}c_{i-\sigma}^{\dagger}+c_{i\sigma}^{\dagger}c_{i-\sigma}^{\dagger}\right)\exp\left(i\mathbf{Q}_{0}\cdot\mathbf{R}_{ij}\right)$
going from 
\begin{align}
 & |\mbox{SC}\rangle=\sum_{\sigma}c_{i\sigma}c_{j-\sigma}|0\rangle\label{eq:0}\\
\mbox{to } & \quad|\mbox{CO}\rangle=1/2\sum_{\sigma}\left(c_{i\sigma}^{\dagger}c_{j\sigma}+c_{j\sigma}^{\dagger}c_{i\sigma}\right)\exp\left(i\mathbf{Q}_{0}\cdot\mathbf{R}_{ij}\right),\nonumber 
\end{align}
where the indices {$i$} and {$j$} refer typically to first nearest
neighbors, $\mathbf{Q}_{0}$ is the modulation wave vector, and $\mathbf{R}_{ij}=\left(r_{i}+r_{j}\right)/2$.
We then have $\eta^{\dagger}|SC\rangle=|CO\rangle$. The rotation
operators between the two states form a SU(2) group with a triplet
representation. At half filling for the Hubbard model, this symmetry
is exact with $\mathbf{Q}_{0}=\left(\pi,\pi\right)$~\cite{Zhang1990}.
Our idea was to generalize this emergent symmetry concept away from
half filling with $\mathbf{Q}_{0}$ being the ordering wave vector
observed in the underdoped region of the phase diagram (see, e.g.,
\cite{Montiel:2016it,Pepin14,Pepin19}).}

{}

\begin{figure}
{\includegraphics[width=8cm]{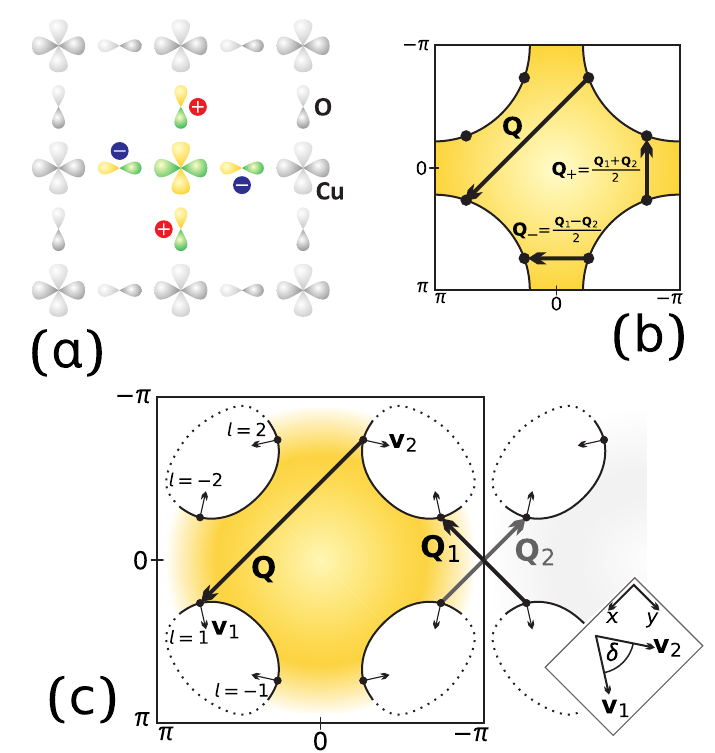} \caption{\label{fig:Real-space-} {(a)} Real space $\mbox{CuO}_{2}$ plane.
{(b) and (c)} Fermi surface with AF {wave} vectors showing the
{location} of the eight hot-spots. From Ref.~\cite{Efetov13}.}
}
\end{figure}

{With K. B. Efetov, we found a model where the emergent
SU(2) symmetry is exact \cite{Efetov13}. It consists of eight hot
spots situated at the AF zone boundary of the Brillouin zone, as depicted
in Fig. \ref{fig:Real-space-}~\cite{Metlitski10,Metlitski10b},
where we see that such a model corresponds to a drastically simplified
version of the Fermi surface, allowing additional symmetries to be
present. Within this restriction of the Fermi surface, we could work
out the model exactly, and found that the two kinds of pairings $\Delta_{pp}=\left\langle \left(i\sigma_{2}\right)_{\alpha\beta}c_{\mathbf{p}\alpha}^{\dagger}c_{-\mathbf{p}\beta}^{\dagger}\right\rangle $
and $\Delta_{ph}=\left\langle c_{\mathbf{p}\alpha}^{\dagger}c_{-\mathbf{p}\beta}\right\rangle $
in the particle-particle and particle-hole channels, respectively,
are fully degenerate. The effective field theory is a non-linear $\sigma$
model with the effective action 
\begin{align}
 & F=\lambda\int d^{2}r\,\text{Tr}\left[\partial u^{\dagger}\partial u+\kappa^{2}u^{\dagger}\tau_{3}u\tau_{3}\right],\label{eq:1}\\
 & \mbox{with }u=\left[\begin{array}{cc}
\Delta_{ph} & \Delta_{pp}\\
\Delta_{pp}^{*} & \Delta_{ph}
\end{array}\right]\nonumber \\
 & \mbox{and }u^{\dagger}u=1,\nonumber 
\end{align}
where $\Delta_{ph}$ is the field order parameter for the particle-hole
pair forming charge density wave, and $\Delta_{pp}$ is the field
order parameter for the particle-particle pairs forming the SC state.
The model in Eq. (\ref{eq:1}) has properties of chirality: the constraint
has to be understood as a local constraint. A study of the eight hot
spots model within a two loop renormalization group has given the
same conclusion of an emergent SU(2) symmetry of the order parameters
~\cite{Freire:2015kg}. This has several experimental consequences:
First, the constraint $u^{\dagger}u=1$, when coupled to fermions,
opens a PG in the antinodal region of the Brillouin zone (see, e.g.,
\cite{Grandadam21}). This situation will be treated in detail later
in this paper, when we study the special case of Bi2201 \cite{Grandadam20,Grandadam21}.
Additionally, typically chiral models like Eq. (\ref{eq:1}) lead
to phase separation. We have treated this idea in a set of works where
the concept of ``droplets'' was introduced to account for the fluctuations
below the PG temperature $T^{*}$ \cite{Meier14,Montiel16,Montiel:2017gf,Montiel:2017gf,Kloss:2016hu}.
}

{WMontiel:2017gfe now turn to the interpretation of the modulations
in the vortex core within the context of the emergent SU(2) symmetry.}

\subsection{{Modulations in the vortex core}}

\begin{figure}
{\includegraphics[width=4cm]{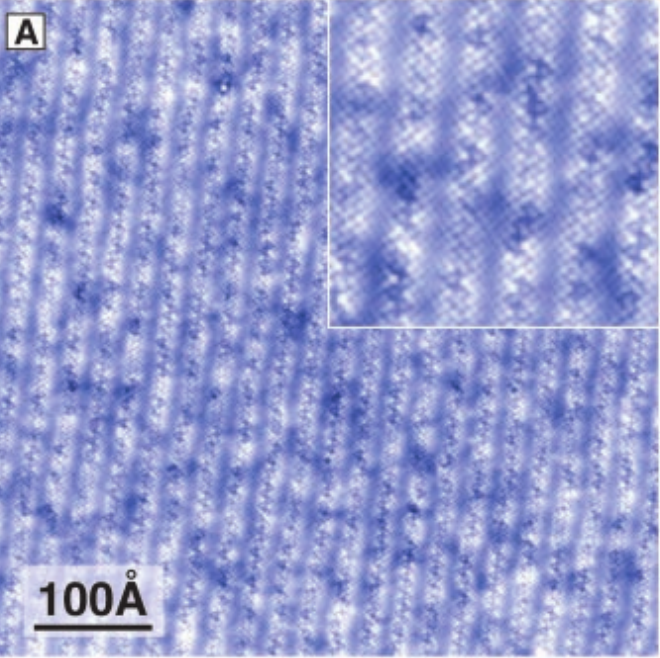}\includegraphics[width=4cm]{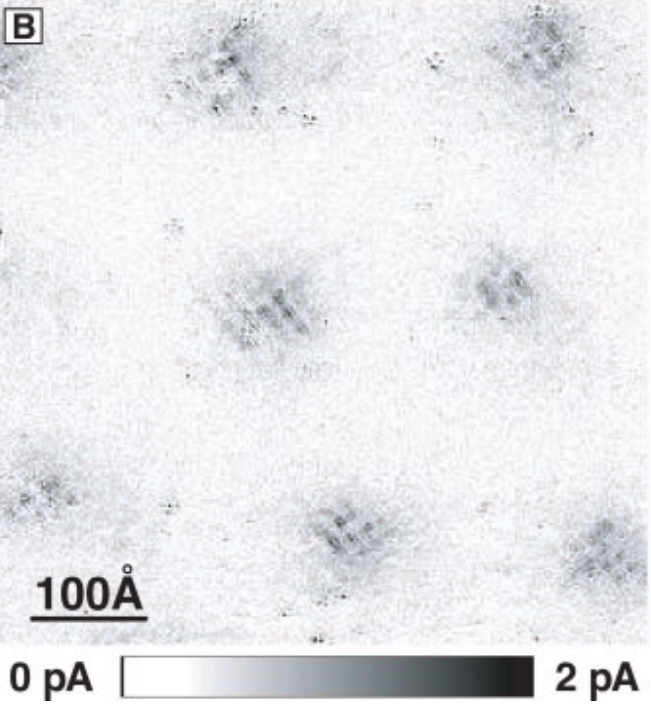}
\caption{\label{fig:Observation-of-the}Observation of the charge modulation
{inside} the vortex core through {scanning tunneling microscopy}
experiment ~\cite{Chakraborty19}. The data in {(a)} are taken
at {$B=0$ T,} whereas in {(b)} the data from {$B=0$ T} and
$B=8$ {T} are subtracted, showing the presence of the {vortex}
lattice at $8$ {T} and the modulations in the core. }
}
\end{figure}

{STM was probably the first experiment that showed
unambiguously the presence of charge order in the underdoped regime
of the cuprates~\cite{Hoffman02}. Measurements in Bi2212 around
12\% of oxygen doping, using an applied field of 8 T and also at zero
field~\cite{Hamidian15,Hamidian15a}, are shown in Fig. \ref{fig:Observation-of-the}.
When the two sets of data are subtracted form each other, the vortex
lattice becomes visible, but also inside the vortex core some modulations
are observed with a period around 3 to 4 lattice sites. The presence
of modulations inside the vortex is a very striking feature which
is not necessarily present for systems with coexistence of CDW and
SC states. In the example of NiSe$_{2}$, where a charge order with
ordering temperature five times as big as the SC temperature is observed,
no modulations are observed inside the vortices~\cite{ForroTiSe209}.
The theory of Eq. (\ref{eq:1}) gives a very elegant explanation for
this feature, which identifies the charge modulations inside the vortex
core to a skyrmion. In order to understand this, we notice that within
the framework of the non-linear $\sigma$ model, a space of pseudospins
is created which mimics the spin operators. }

\begin{figure}
{\includegraphics[width=8cm]{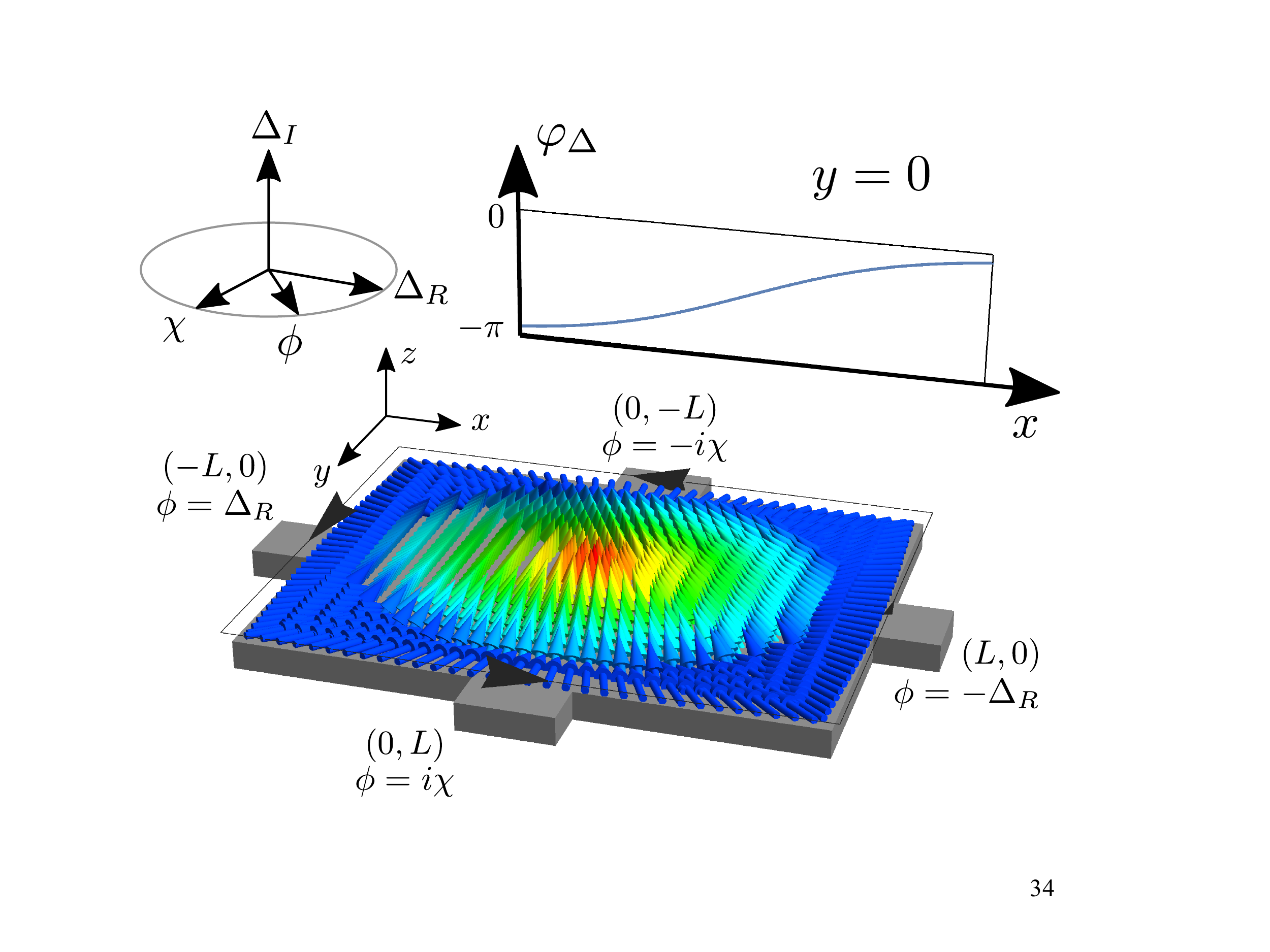} \caption{\label{Fig5}Depiction of a skyrmion in the {pseudospin} space.
Within the framework of emergent symmetry, the skyrmion {agrees with}
the experimental observation of {a} modulation inside the vortex
core. We see that the {pseudospin} goes from a position aligned
with the SC order parameter to a position aligned with the CDW inside
the vortex.}
}
\end{figure}

{The correspondence goes as 
\begin{align}
S_{x} & \rightarrowtail\Delta,\nonumber \\
S_{y} & \rightarrowtail\Delta^{*},\nonumber \\
S_{z} & \rightarrowtail\chi,\label{eq:2}
\end{align}
and {vector} rotations operate between the three components, which
are depicted in Fig. \ref{Fig5}. Away from the vortex core{,} the
vector is aligned with the SC state, say{,} in the {$x$}-direction.
Going inside the vortex core, the vector turns progressively toward
the {$z$}-direction, reaching precisely the {$z$}-direction
inside the core. The result is a skyrmion {in} the space of {pseudospins}~\cite{Pepin14,Montiel16,Kloss:2016hu,Kloss15,Einenkel14}.
Note that this feature that {a} chiral model {produces} skyrmions
is general was one of the {predictions} of the theory of emergent
symmetries in the past. For example, within the SO(5) theory, the
competitor state being AF order, it was predicted that AF order would
be seen inside the vortex core. Similarly, within the SU(2) gauge
theory of Ref. \cite{Lee06}, a state with loop currents was predicted
inside the vortex core. The SU(2) theory which rotates from {a}
SC state to {a} CO state elegantly satisfies the prediction that
charge modulations would be seen inside the vortex core. So the theory
passes this experimental test. However, we note that the test is not
decisive, since it has been argued that a simple, but strong enough
Ginzburg{-}Landau (GL) interaction between the two modes is enough
to create charge modulations inside the vortex core (see, e.g., \cite{Fradkin15}).
}

\subsection{{B-T phase diagram}}

\begin{figure}
{\includegraphics[width=8cm]{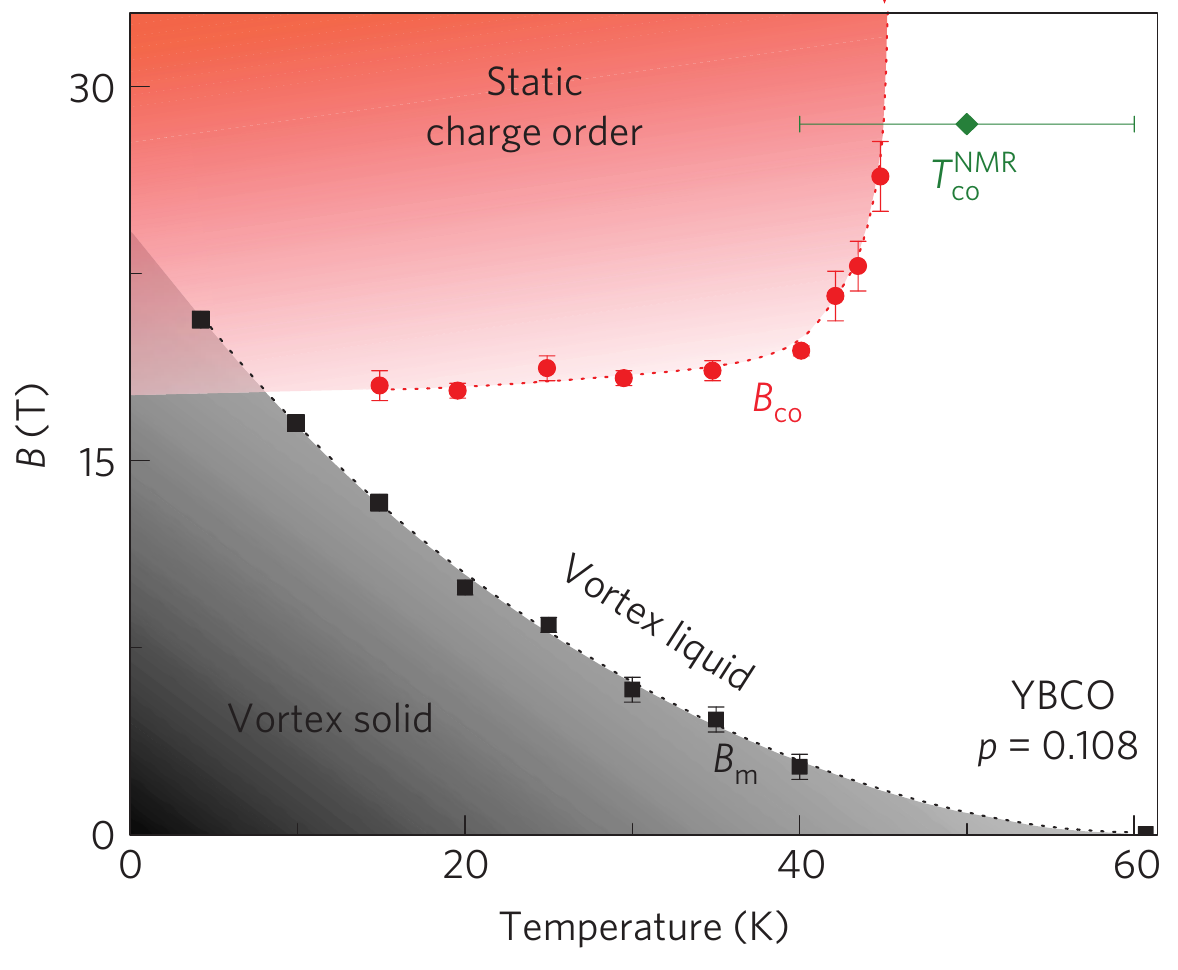}}

{\includegraphics[width=8cm]{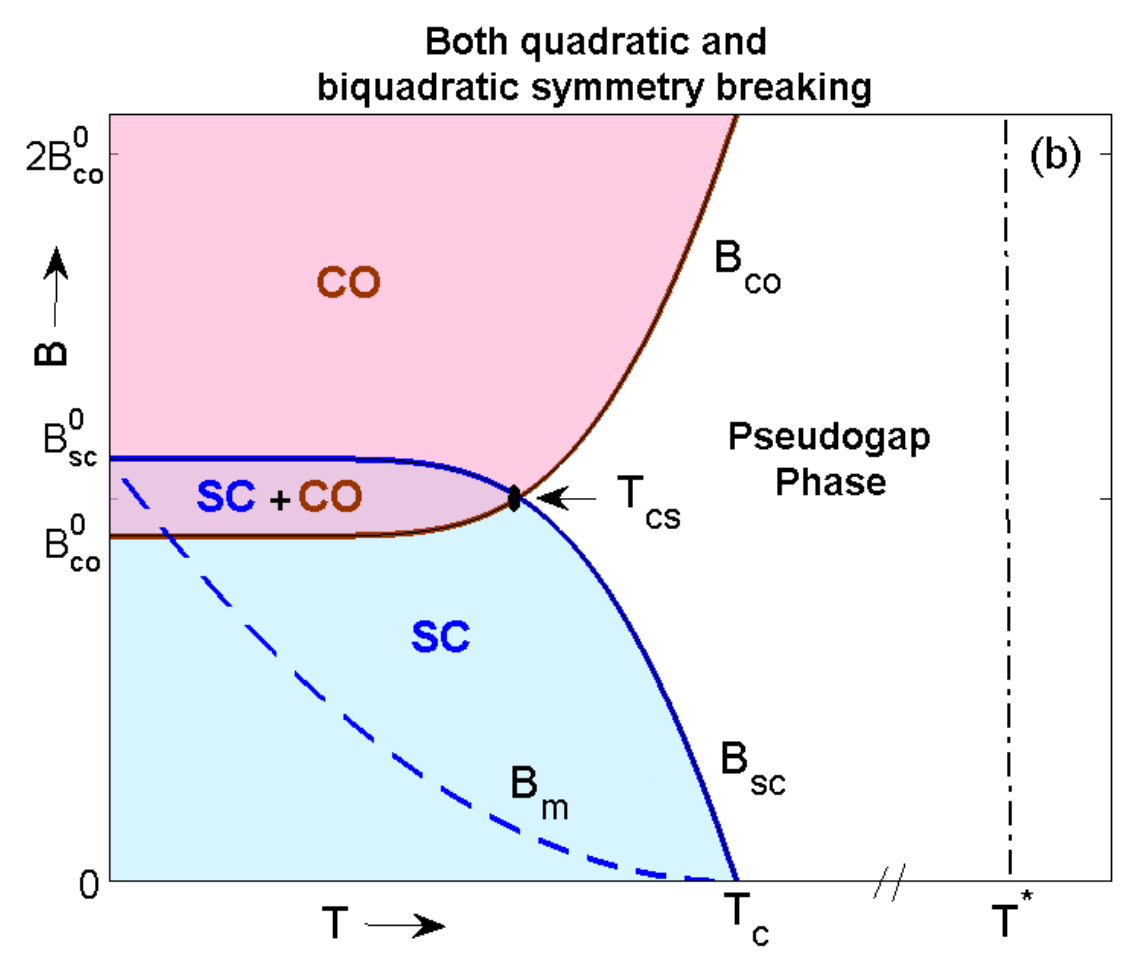}}

{\caption{\label{Fig6}{(Upper panel)} The experimental {$B$-$T$} phase
diagram{,} where the flatness of the transition with respect to
the temperature can be observed~\cite{LeBoeuf13}. {(Lower panel)}
The simulation with the {non-linear $\sigma$} model~\cite{Chakraborty18}. }
}
\end{figure}

{We now know that SC and CO orders in {underdoped}
cuprates interact strongly {with each other}, but what is the form
of their interaction{?} {They} could do it through a simple GL
coupling $\lambda\left|\Delta_{i}\right|^{2}\left|\chi_{i}\right|^{2}$,
where $\Delta_{i}$ and $\chi_{i}$ are{,} respectively{,} the
SC and CO order parameters, but they could as well be ``entangled''
and interact through a local constraint $\left|\Delta_{i}\right|^{2}+\left|\chi_{i}\right|^{2}=1$.
In both cases{,} the two orders are in competition, but in the second
case, the constraint produces an entangled ground state with the two
orders. The issue here {is} whether the effective model describing
the PG phase is chiral or not chiral. In order to examine the situation,
we turn to a very powerful set of experiments where the phase diagram
in the {$B$-$T$} plane was determined ~\cite{LeBoeuf13,LeBoeuf11,Chang:2016gz},
and depicted in Fig. \ref{Fig6}. The compound studied through {ultrasound}
and x-ray spectroscopy is YBCO at 12\% of oxygen doping. As shown
in Fig. \ref{Fig6}, the phase diagram shows a SC phase at $B=0$
up to a temperature $T_{c}\sim45$ {K}. Above a magnetic field of
$B_{c}\sim17$ T{,} the CO sets in {abruptly}, with a very flat
phase transition and a transition temperature of the same magnitude
as $T_{c}$, with $T_{co}\simeq T_{c}$. The flatness in temperature
of this phase transition is {one} of the most striking features
of the {$B$-$T$} phase diagram. It is reminiscent of a spin-flop
transition in magnetic structures when the magnetic field is applied
perpendicular to the orientation of the spin{:} nothing happens
for a while when the magnitude of the magnetic field is increased,
but suddenly all the spins {``flip''} to get aligned to the magnetic
field, producing a very abrupt and flat phase transition.}

{We have investigated the {$B$-$T$} phase diagram
in two papers~\cite{Chakraborty18,Meier13}. As could be expected
from the arguments above, the chiral models describe naturally the
spin flop transition whereas the simple GL theory has to be heavily
{fine-tuned} to account for the observed features. At low coupling
between the two modes, the GL theory cannot produce a flat transition{:}
the coupling has to be ``strong'', typically larger than one in
renormalized units {to exhibit} the flat transition. Now, if the
coupling is larger than one, the two orders repel each other fully
and it is impossible to get a {coexisting} phase. This is the difference
with the ``entanglement'' between competitive {orders} given by
chiral models. The constraint of chiral models in Eq. (\ref{eq:1})
naturally produces a flat ``spin-flop'' transition, but in the space
of {pseudospin}. In order to account for the {coexistence}, the
exact SU(2) symmetry has to be {slightly} broken, by adding for
example a mass term in Eq. (\ref{eq:1}), but there are many other
ways to break the exact SU(2) symmetry{,} as we will see a bit later.
The conclusion of our study of the {$B$-$T$} phase diagram is
that the chiral model accounts for the experimental {data} one order
of magnitude better than the GL theory. Note that a chiral model (the
non-linear $\sigma$ model) has been numerically studied by other
authors to describe the interplay of SC and CDW with impurities ~\cite{Caplan15,Caplan17}.
}

\subsection{{Raman scattering and charge gap\label{subsec:Raman-scattering-and}}}

\begin{figure}
{\includegraphics[width=8cm]{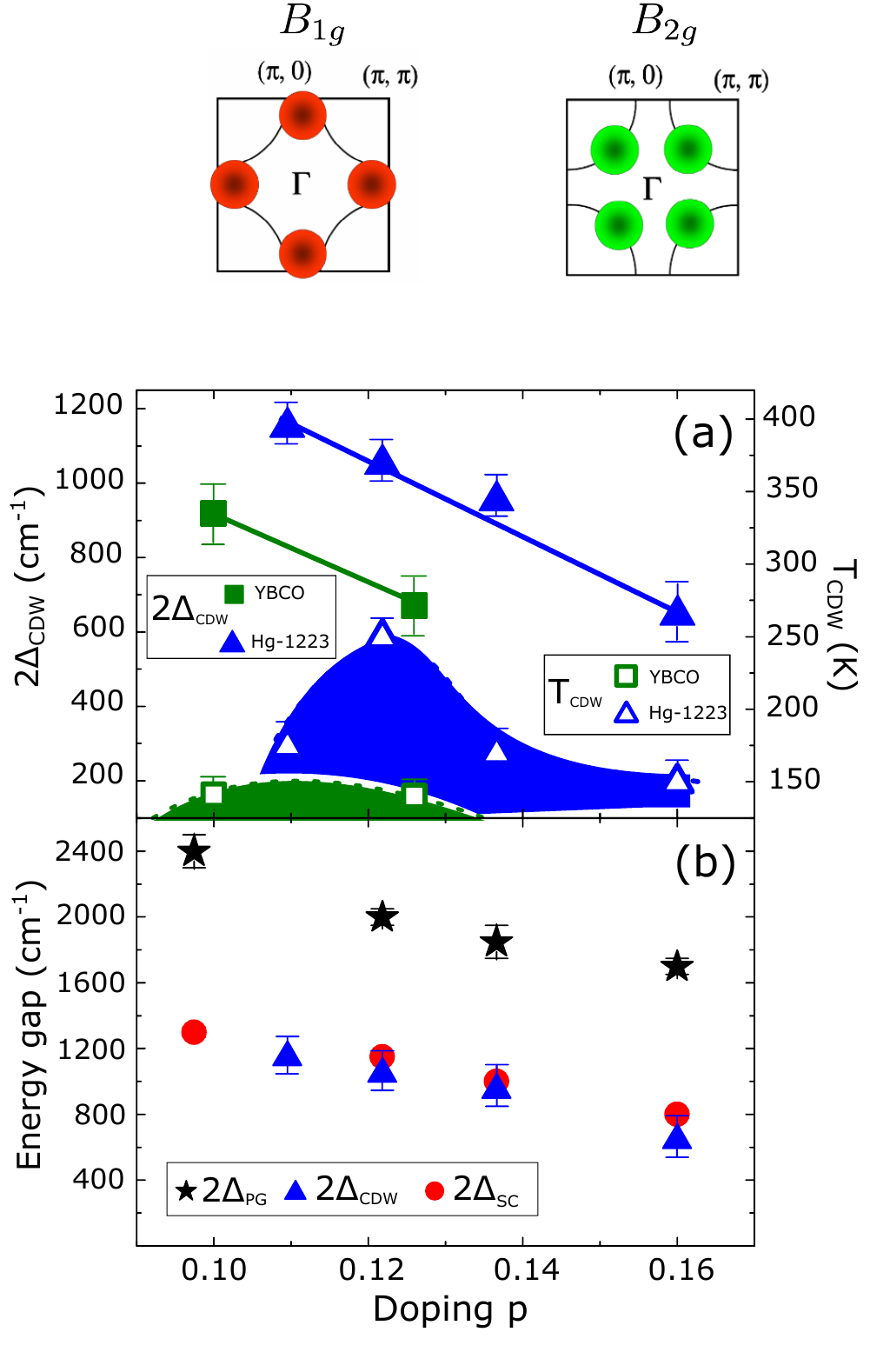} \caption{\label{Fig7} Raman $B_{1g}$ and $B_{2g}$ channels scan{, respectively,
the antinodal} and the nodal parts of the Brillouin zone. {The charge}
gap is extracted from the $B_{2g}$ channel. We observe that the charge
gap is of the same order of magnitude as the SC gap, and varies like
$T^{*}$ (and not {like} $T_{co}$) with doping. From Ref.~\cite{Loret19}.}
}
\end{figure}

{A question very much debated with the observation
of charge order in {underdoped} cuprates is whether it is a key
stone for our understanding of the remaining mysterious phases (as
we advocate in this paper), like the PG, or whether it is an {epiphenomenon,}
which comes as a decoration of the phase diagram but is not central
to it. A very intriguing experiment from Raman scattering claims to
be able to extract the amplitude of the charge gap from the $B_{2g}$
resonance~\cite{Loret19}. As a reminder{,} the Raman spectroscopy
is able to scan parts of the Brillouin zone (BZ) by selecting different
symmetries as depicted in the upper right panel of Fig. \ref{Fig7}.
The $A_{1g}$ channel is uniform around the BZ and thus {does not}
select any symmetry. The $B_{1g}$ channel scans the {antinodal}
(AN) region of the BZ{,} whereas the $B_{2g}$ channel scans the
nodal region roughly up to the AF zone boundary. When entering the
SC phase{,} one can see in Fig. \ref{Fig7} a {peak} is seen in
Raman scattering at $\omega=2\Delta$, both in the $B_{1g}$ and $B_{2g}$
channels. In the $B_{1g}$ channel, however, we have an additional
feature visible as a ``{peak}-dip-hump'' (the same feature is
also visible in momentum resolved ARPES around the same part of the
BZ (see, e.g., \cite{Vishik:2010fn} and references therein), which
is identified as a signature of the PG line. The novelty of Raman
scattering studies is that a ``new'' {peak} is observed in the
$B_{2g}$ channel, which is separate from the SC {peak}. This has
been identified with the CO gap in the {underdoped} region. This
charge gap has two remarkable features: {First,} it is of the same
order of magnitude as the SC gap, which provides an answer that it
cannot be an epiphenomenon. Second, the gap varies with doping, but
{does not} follow the CO transition temperature $T_{co}$. Instead,
it follows the PG temperature $T^{*}$.}

{Those two ingredients are part of the SU(2) emergent
symmetry scenario, since in this case, the two gaps {(the charge
gap $\chi$ and the SC gap $\Delta$)} behave essentially in the
same way, {i.e., they have} the same order of magnitude and the
same variation with doping ~\cite{Efetov13}. One consequence of
the Raman experiment is that at {low} doping the charge sector has
a ``preformed gap'' in the same manner as the SC sector. Namely{,}
the bare value of the gap, both in the charge and SC sectors{,}
is larger than the ordering temperature {$T_{co}$}. There is a
fluctuation regime where the gap is {``pre-formed'',} but not
condensed in both the charge and the SC sectors~\cite{Grandadam19}.
This last remark adds further support to the idea that strong fluctuation
regime in the underdoped region is an important key to understand
the PG phenomenon. }

\subsection{{Inelastic Neutron Spectroscopy (INS) and $\eta$
mode}}

\begin{figure}
{\includegraphics[width=8cm]{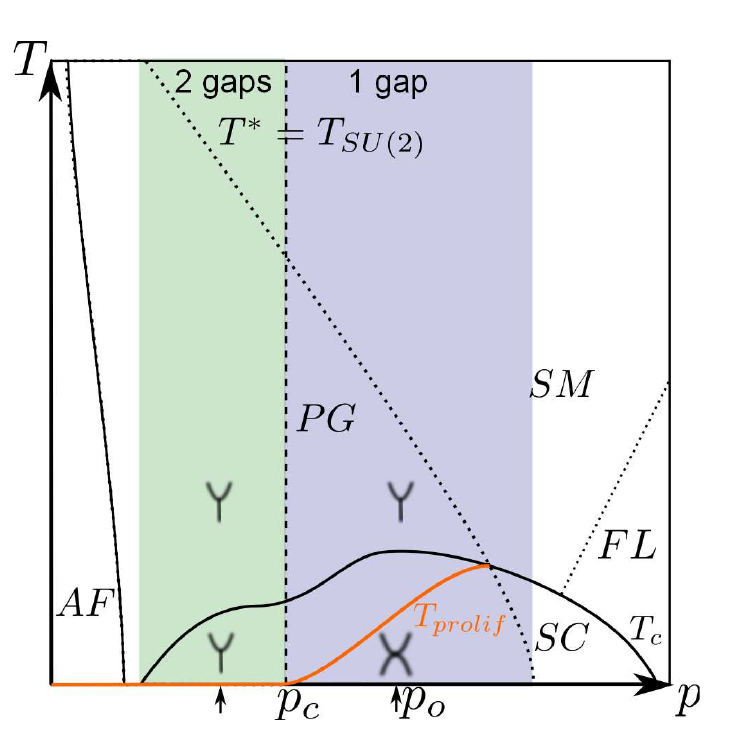} \caption{\label{fig:Schematic-phase-diagram}Schematic phase diagram showing
the various type of neutron resonances in the SC phase and PG phases
of {the cuprate} superconductors. From Ref. \cite{Montiel17}.}
}
\end{figure}

{The idea of emergent SU(2) symmetry {has been}
criticized from a few standpoints. The most important criticism is
whether the symmetry survives the full Fermi surface (see, e.g.,
the discussion in Ref. \cite{Fradkin15}). Indeed, so far the SU(2)
symmetry was found to be exact on the eight-hot-{spot} model which
consists of eight points situated at intersection of the AF zone boundary
and the Fermi surface. But what happens when the whole Fermi surface
is involved{?} {Will} the symmetry survive, even for a point of
the Fermi surface not far way from the eight hot spots{?} In order
to address this criticism, we first considered a model of ``droplets''
made with charge modulations involving a series of wave vectors around
the original CO one $\mathbf{Q}_{0}$~\cite{Montiel:2017gf,Kloss:2016hu,Pepin14}.
This idea pushed us to consider ``hot regions'' of the Fermi surface,
situated in the AN part of the BZ. The INS experiment tells us that
there is a ``butterfly shape'' of the neutron scattering resonance
situated at the vector $\mathbf{Q}=\left(\pi,\pi\right)${,} but
at {a} finite energy inside the SC phase. The same resonance changes
to a ``Y'' shape between $T_{c}$ and $T^{*}$, as depicted in the
purple region of the phase diagram in Fig. \ref{fig:Schematic-phase-diagram}~\cite{Hinkov04,Hinkov07}.
At lower oxygen doping below $12\%$, the ``Y'' shape is seen ubiquitously
below and above $T_{c}$, as depicted in the green region in Fig.
\ref{fig:Schematic-phase-diagram}. The standard interpretation of
the butterfly resonance below $T_{c}$ is that the upper part is the
dispersion of AF spin density waves. It is gapless at zero doping
and gradually becomes gapped at finite doping~\cite{Eremin:2005ba}.
The origin of the lower part has led to many controversial ideas,
with the most credibility given to the scenario of magnetic excitons
showing below $T_{c}$~\cite{Norman:2001vfa,Norman:2007eq,NormanChub01}.
Within the exciton scenario, the shape of {the} lower part of the
``butterfly'' is driven by the shape of the {$d$}-wave dispersion
of the gap around the node, a fact that is well{-}reproduced experimentally.
Although the ``butterfly'' shape is rather well{-}understood theoretically,
the change into a ``Y'' shape above $T_{c}$ resists interpretation.
Within our scenario of ``hot regions'' of the Fermi surface, we
interpret this result with the idea of a loss of coherence around
the AN region of the Fermi surface when the $T_{c}$ line is crossed.
With this idea{,} we were able to reproduce the entire phase diagram
for the INS resonance~\cite{Montiel17}, as depicted in Fig. \ref{fig:Schematic-phase-diagram}.
}

{Another important ingredient of emergent symmetries
is that they are accompanied by collective modes. In the case for
SO(5) group described earlier for a rotation between AF and SC ground
states, the collective mode was considered as the explanation for
the neutron scattering resonance \cite{Demler95}. For the SU(2) symmetry,
the collective mode $\eta$ has been studied in order to explain the
mode seen in the $A_{1g}$ channel in Raman scattering spectroscopy
\cite{SacutoSidis} in a paper where the dependence on doping is contrasted
with the dependence on doping of the INS resonance \cite{Montiel15a}.
The collective mode has also been thought for explaining the incongruous
dispersive feature \cite{Chaix:2017fs} observed by x-ray scattering
around the charge ordering wave vector $\mathbf{Q}_{0}$ \cite{Morice18b}.}

\section{{Fractionalized Pair Density Wave\label{sec:Fractionalized-Pair-Density}}}

\subsection{{The concept}}

{We then turned to a more drastic solution to the
criticism, with a new chiral model that encompasses the same constraint
as in Eq. (\ref{eq:1}), but which sees an emergent U(1) gauge field
be responsible for the formation of the PG. The main context is the
formation of hard core bosons at high energy (of the order of 0.5
eV) in the phase diagram of the cuprates. This is our starting hypothesis.
These bosons may have different symmetries (singlet, triplet, charge
zero or charge two) and, due to their hopping from site to site, a
continuum of wave vectors is explored. Of course{,} this idea is
difficult to prove theoretically, but we will see later that some
experimental evidence inclines towards it.}

{Out of this {boson} ``soup'' which interacts
with the fermions, some bosons will condense at low energy (forming
the various orderings seen in the {underdoped regime}), but {also}
at some {places in the overdoped regime} in the phase {diagram}
it could be the result of such a condensation ~\cite{Chakraborty19,Grandadam19}.
On the other hand{,} charge two particle-particle bosons with finite
center{-}of{-}mass {momentum}, also called {pair-density waves}
(PDW), might be unstable when {the} temperature is lowered and ``fractionalize''
into their elementary symmetry components $\hat{\Delta}_{{PDW}}=\left[\hat{\Delta},\hat{\chi}\right]$,
where $\hat{\Delta}=\left\langle \sum_{\sigma}c_{i\sigma}c_{j-\sigma}\right\rangle $
is {a} charge-two, $\mathbf{Q}=0$ particle-particle pair and $\hat{\chi}=\left\langle \sum_{\sigma}c_{i\sigma}^{\dagger}c_{j\sigma}e^{i\mathbf{q}\cdot\mathbf{R}_{ij}}\right\rangle $
is the particle-particle order breaking the translation symmetry,
with $\mathbf{R}_{ij}=\left(\mathbf{r}_{i}+\mathbf{r}_{j}\right)/2$.
The system has the same local constraint as Eq. (\ref{eq:1}) with
\begin{align}
\left|\Delta_{R}\right|^{2}+\left|\chi_{R}\right|^{2} & =\left(E^{*}\right)^{2},\label{eq:4}
\end{align}
where $E^{*}$ is a constant (note that both $\Delta_{R}$ and $\chi_{R}$
have dimension of energy). The effective model is now another {chiral
model (i.e., a model controlled by a local constraint),} which can
be written in the form of a quantum rotor model  (see section 3 in Ref.~\cite{Perelomov81}) :
\begin{align}
 & S=\int d^{2}x\sum_{a,b=1}^{2}\left|\omega_{ab}^{\mu}\right|^2,\nonumber \\
 & \mbox{with }\omega_{ab}^{\mu}=z_{a}\partial_{\mu}z_{b}-z_{b}\partial_{\mu}z_{a},\nonumber \\
 & \mbox{and }\sum_{a=1}^{2}\left|z_{a}\right|^{2}=1,\label{eq:5}
\end{align}
with $z_{1}=\Delta_{R}/E^{*}$, $z_{2}=\chi_{R}/E^{*}$, $z_{1}^{*}=\Delta_{R}^{*}/E^{*}$,
$z_{2}^{*}=\chi_{R}^{*}/E^{*}$. The model has a natural U(1) gauge
symmetry {with} $z_{a}\rightarrow z_{a}e^{i\theta}$ ($z_{a}^{*}\rightarrow z_{a}^{*}e^{-i\theta}$).
The model {in} Eq. (\ref{eq:5}) is formally equivalent to the $\mbox{CP}^{1}$
model 
\begin{align}
 & S=\int d^{2}x\left|D_{\mu}\psi\right|^{2},\label{eq:6}\\
 & \mbox{with }D_{\mu}=\partial_{\mu}-i\alpha_{\mu},\nonumber \\
 & \mbox{and }\alpha_{\mu}=\frac{1}{2}\sum_{a}z_{a}^{*}\partial_{\mu}z_{a}-z_{a}\partial_{\mu}z_{a}^{*},\nonumber 
\end{align}
with $\psi=\left(z_{1},z_{2}\right)^{T}$. The model in Eq. (\ref{eq:6})
is in turn ``almost'' the same as the {non-linear $\sigma$} model
of Eq. (\ref{eq:1}){,} but with an additional gauge field $\alpha$
taking care of the intrinsic U(1) gauge symmetry.}

{An advantage of this new formulation is that the
constraint {in} Eq. (\ref{eq:4}) is now very robust because {it
is} protected by an emergent gauge field. Within this new formalism,
the PG of the phase diagram of the cuprates can be interpreted in
the following way. At high temperature{,} PDW bosons are present
in the {``boson soup''}. When the temperature is decreased, as
depicted in Fig. \ref{fig:Phase-diagram-of}, we observe a deconfining
transition at $T^{*}$ followed by a {reconfining} transition at
lower temperature when each of the elementary field ($\Delta$ and
$\chi$) is condensed. Note that this model is not the only one with
a fractionalization of a composite ``boson''. An attempt at describing
the phase diagram was made through ``stripe fractionalization'',
using a discrete $Z_{2}$ emergent gauge symmetry \cite{Zaanenstripes}.
The present idea has similarities with the concept of emergent gauge
theories at optimal doping ~\cite{Sachdev19}, or those of a fractionalized
Fermi liquid \cite{Senthil:2001jm,Senthil03}, but differs in the
sense that the fractionalization of a PDW or a stripe does not affect
the elementary particles. It is closer to the concept of entanglement
of two elementary fields (here, $\chi$ and $\Delta$) subject to
a local constraint, which makes the whole system equivalent to a ``pseudo-spin''
space.}

\subsection{{Angle-Resolved Photoemission Spectroscopy (ARPES) }}

\begin{figure}
{\includegraphics[width=9cm]{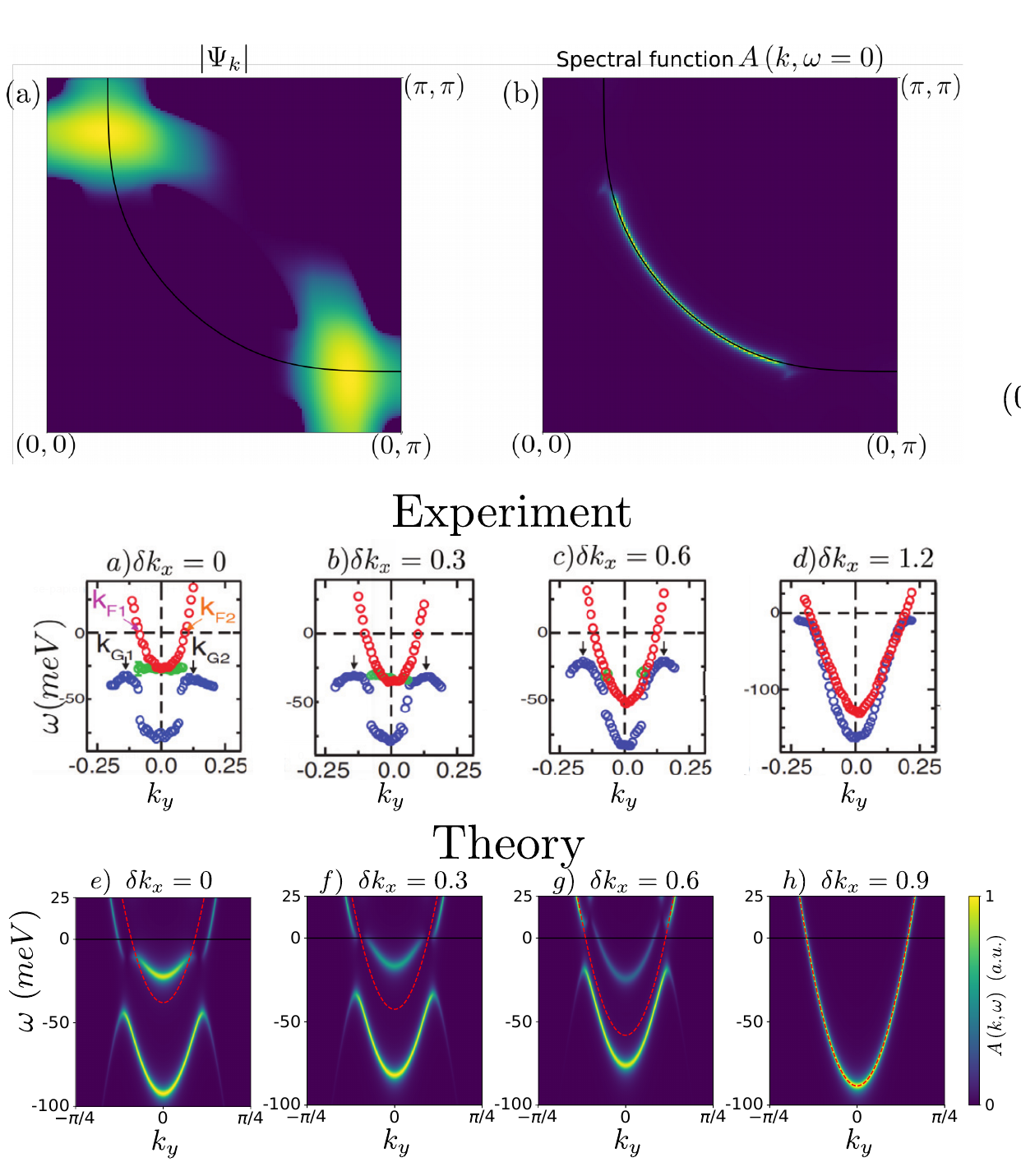} \caption{\label{fig:The-opening-of}The opening of the gap in the AN region
of the Brillouin zone is shown above. The {experimental} data on
$\mbox{Bi2201}$ from Ref. ~\cite{He11} and the simulation of the
same data within our model~\cite{Grandadam20,Grandadam21}.}
}
\end{figure}

{Now that our idea is generalized to the whole Fermi
surface (and not only restricted to the eight-hot-spot model) we can
study the ARPES experiment in order to determine, in particular, how
the Fermi surface is affected by CO and SC gaps, away from the eight
hot-spots~\cite{Grandadam20,Grandadam21}. The first check that we
have done is to see that the constraint {in} Eq. (\ref{eq:4}),
when coupled back to the fermions in a mean-field theory model, opens
a gap in the AN region of the Fermi surface, precisely in the same
region that was identified by ARPES as the PG region (see, e.g., \cite{Vishik18}).
This is shown in the upper part of Fig. \ref{fig:The-opening-of},
where the gap regions are shown in panel $(a)$ and the spectral density
is shown in panel $(b)$. This leads to the formation of the so-called
``Fermi arcs'' around the nodal region of the Fermi surface. The
particularity of the opening of the gap on part of the Fermi surface
is due to the wave vector of the CO gap, which affects only part of
the {conduction} electrons, typically in the AN region of the Fermi
surface. Moreover, this scenario belongs to the category of ``fluctuation''
theories, which {were} intensively studied in the past in the context
of the ARPES data, considering the fluctuations of the phase of the
pairing order parameter $\Delta$ ~\cite{Kanigel:2008wm}. In our
scenario{,} both the phase and the amplitude of the two order parameters
$\Delta$ and $\chi$ vary, only related by the constraint {in}
Eq. (\ref{eq:4}). One very nice feature of the fluctuation scenarios
compared to the other ones is that the PG gap is naturally closing
with temperature at $T^{*}$, and the {Fermi} ``arcs'' are getting
smaller and smaller as the temperature is increased (finally vanishing
at $T^{*}$), which lead to a large Fermi surface observed in experiments.
It {should} be mentioned that, among the various scenarios for the
formation of the PG, the fluctuation scenario is the only one which
reproduces the very gradual closing of the {Fermi} surface with
the temperature. It is also important to notice that so far only theories
based on a fluctuation scenario account for the closing of the Fermi
arcs at $T^{*}$ observed experimentally ~\cite{Kanigel:2008wm}.}

{We now turn to the case study of Bi2201{,} from
which extensive ARPES experiments have been performed ~\cite{He11}.
The study of this compound has got some visibility because of two
{observations} in the opening of the PG. First{,} the opening
of the PG in the AN of the Fermi surface occurs at specific wave vectors
}\textbf{{$\mathbf{k}_{G}$}}{,
which are different from the Fermi wave vector $\mathbf{k}_{F}$ identified
above $T^{*}$ when the PG closes. Then there is a back-bending of
the electronic dispersion precisely at the point $\mathbf{k}_{G}$
where the gap opens{,} as can be seen in Fig. \ref{fig:The-opening-of}.
These two {observations} actually restrict very strongly the realm
of possibilities for the PG, since the back-bending indicates that
a particle-hole transformation has to be present{,} whereas the
finite wave vector different from $\mathbf{k}_{F}$ is calling for
a breaking of translational invariance. Note that the observed wave
vector $\mathbf{k}_{G}$ corresponds well to a charge ordering of
wave vector $\mathbf{Q}_{0}\simeq0.3$. All these observations led
to the proposal of the concept of Amperean pairing, with a fluctuating
PDW {as} a candidate for the PG \cite{Lee14}. Recently, an approach
with emergent gauge field has also been proposed to account for this
experiment \cite{SachdevBi2201}. Within the fractionalized PDW scenario,
we were able to reproduce the experiments, accounting for both features
of the back-bending of the dispersion at $\mathbf{k}_{G}$ and the
closing of the Fermi arc with temperature around the Fermi surface.
Our results are depicted in the lower part of Fig. \ref{fig:The-opening-of}
for direct comparison with experiments \cite{Grandadam20}. To our
knowledge, those are the three only attempts to describe this experiment,
the three of which require non-trivial field-theoretic ideas. In that
sense, this ARPES experiment is important for unveiling the mystery
of the pseudogap: it requires a theory to account for a finite momentum
scale $\mathbf{k}_{G}$ (different from the Fermi wave vector $\mathbf{k}_{F}$),
which is surprisingly related to the charge ordering wave vector $\mathbf{Q}_{0}=2\mathbf{k}_{G}$.
It also requires a particle-hole transformation, or some other idea
(as described in Ref. \cite{SachdevBi2201}) in order to account for
the back-bending of the quasi-particle dispersion.}

{}

\subsection{{CDW Phase locking from STM experiment}}

\begin{figure}
{\includegraphics[width=8cm]{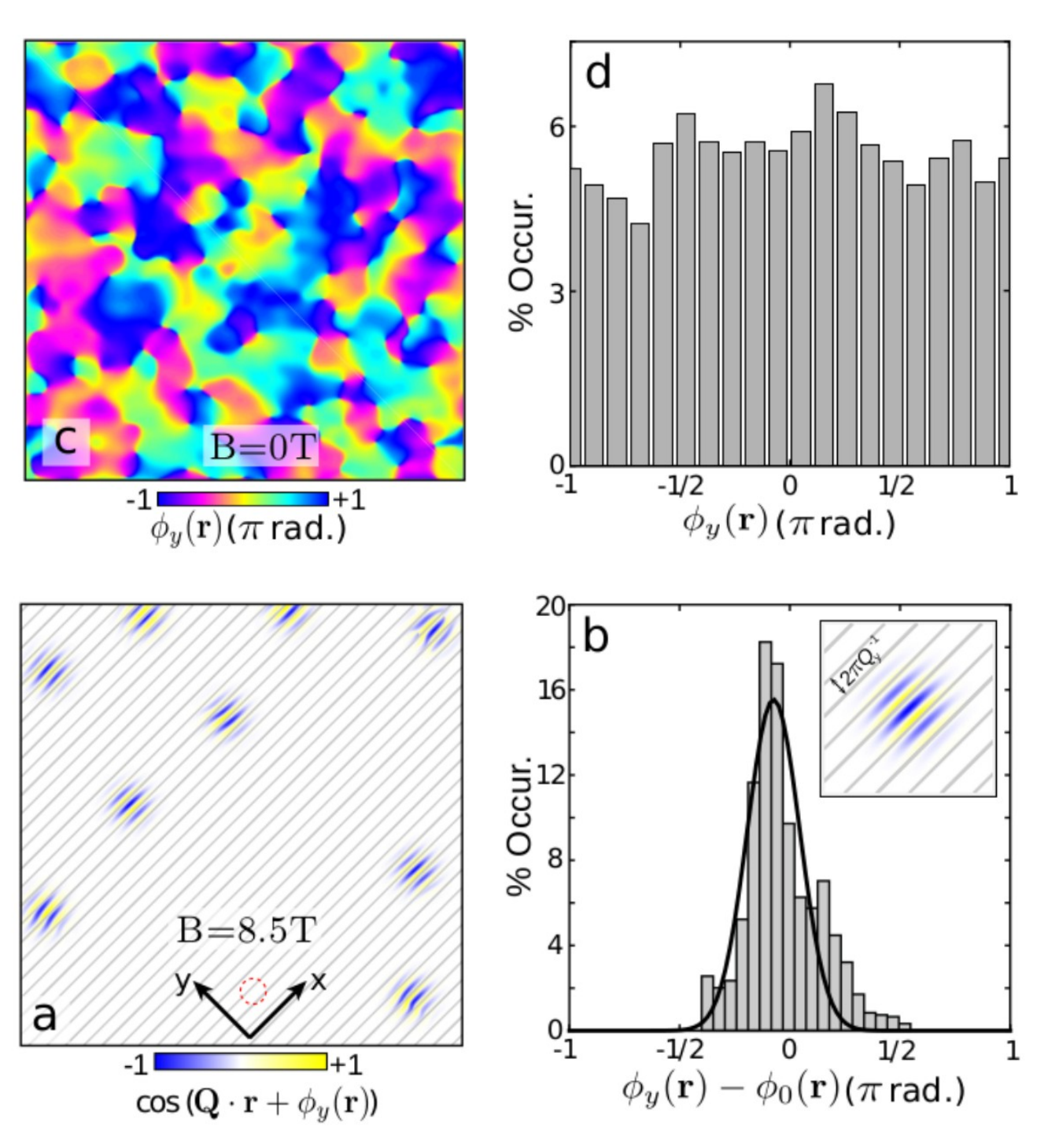} \caption{\label{fig:Above-is-the}{The} STM experiment at $B=0$ {T} showing
a completely random dispersion of the charge order phase $\theta$,
with $\chi=\chi_{0}\cos\left(\mathbf{Q}_{0}\cdot\mathbf{r}+\theta\right)$.
{In the lower panels,} the same study inside the vortex core, when
the data for $B=8.5$ {T} and for $B=0$ {T} have been subtracted.
We {see} that now $\theta$ is \textquotedblleft frozen\textquotedblright{}
around the whole sample{,} whereas the correlations of the amplitude
of the CDW are only running over a few lattice sites. From Ref.~\cite{Chakraborty19}.}
}
\end{figure}

{A very striking experiment, as already mentioned
in the introduction, is the study of the phase slips inside the CO
and SC phases. Recent STM experiments were carried out and were able
to extract the phase $\theta$ in the modulations amplitudes $\chi=\left|\chi_{0}\cos\left(\mathbf{Q}\cdot\mathbf{r}+\theta_{r}\right)\right|$~\cite{Chakraborty19}.
As {can be seen} in Fig. \ref{fig:Above-is-the}, at zero field,
the histogram distribution of $\theta_{r}$ is totally random, spreading
on all values. The surprise comes when the values at $B=8.5$ {T}
is subtracted from the values at $B=0$ {T}. One sees clearly the
vortices with the modulations inside, but then the phase slip $\theta_{r}$
of the modulations is locked at a unique value, {say,} $\theta_r=0$
(see the second panel of Fig. \ref{fig:Above-is-the}). The phase locking
is {long-ranged}, of the size of the sample, whereas the charge
order itself is {short-ranged}, spreading typically about four to
five lattice sites. We are thus {in} a situation where the phase
locking in the charge sector is much {longer} than the correlations
of the amplitude of the order parameter. This is unique to {the}
cuprates. Within the PDW fractionalization scenario, we address this
situation by noticing that the phase of the CDW and the one of the
SC order parameters are related by a gauge phase, and lock together
below $T_{c}$. This has striking consequences, both in terms of the
symmetry of the charge order, which behaves typically like a PDW below
$T_{c}$, and in particular reacts to an {electromagnetic} field
(note that a {reconfining} transition occurs below $T_{c}$). A
second consequence is that the phase $\theta_{r}$ of the charge order
is locked to the SC phase and acquires {long-range} correlations
across the whole sample. }

{This experiment is so difficult to interpret that,
to our knowledge, the interpretation with the ``phase locking''
of the emergent gauge field is to date the only one in the literature.
Note that a more in-depth study of the effect of the locking of the
phases proposed by this approach within a setting of Josephson junctions
has been recently proposed. For the latter study, we refer the reader
to \cite{Banerjee21}.}

\subsection{{Phonon softening}}

\begin{figure}
{\includegraphics[width=8cm]{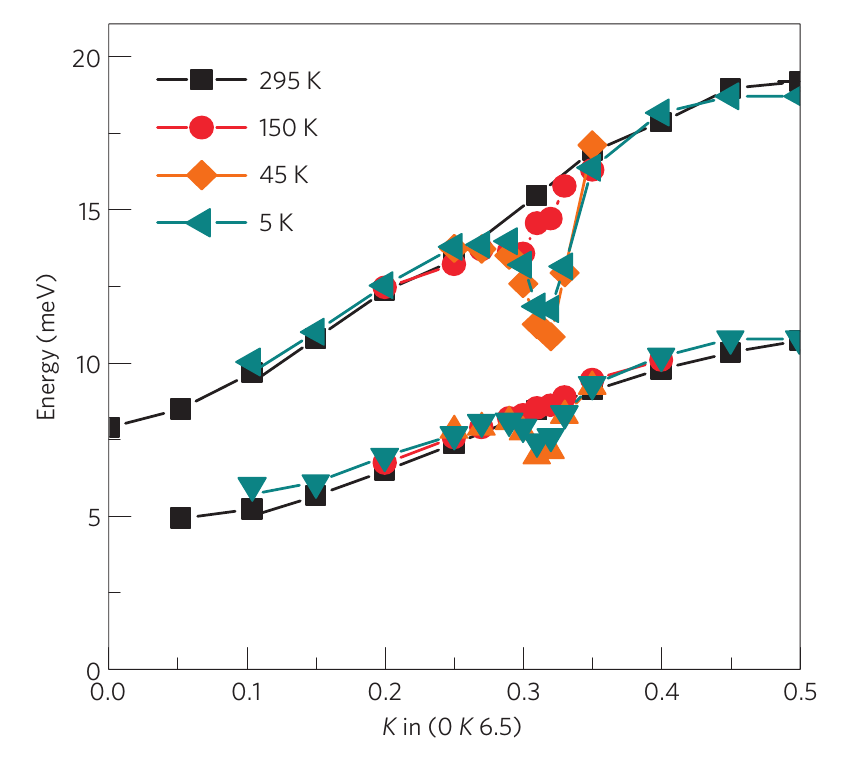} \includegraphics[width=8cm]{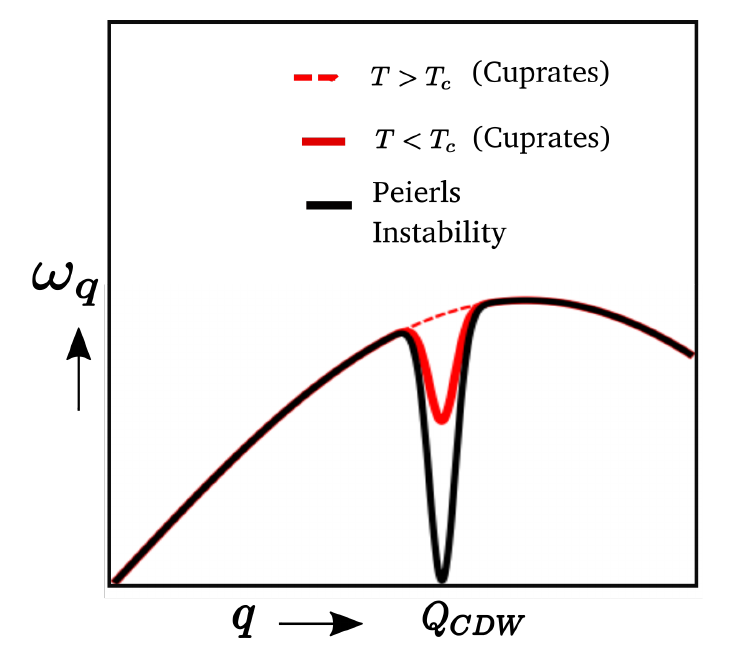}
\caption{\label{fig:Experimental-obervation-of} {(Upper panel)} Experimental
{observation} of the phonon softening below $T_{c}$, from Ref~\cite{LeTacon14}. {(Lower panel)}
The schematic depiction of the phenomenon~\cite{Blackburn13b}.}
}
\end{figure}

{In the same {line} of ideas{,} we considered
the experiment of softening of the phonon line due to the presence
of a CDW order~\cite{LeTacon14,Blackburn13b}. This experimental
result is depicted in Fig. \ref{fig:Experimental-obervation-of}.
The standard Peierls theory of CO stipulates that below the CO transition
temperature $T_{co}$, and at the charge ordering modulation wave
vector $\mathbf{Q}_{0}${,} the phonon line is softened, namely{,}
the phonon dispersion $\varepsilon_{q}\rightarrow0$ at $\mathbf{q}=\mathbf{Q}_{0}$~\cite{Lee79}.
As depicted in the second panel of Fig. \ref{fig:Experimental-obervation-of},
the theory actually works for one dimensional systems. For 2D and
3D systems, a softening is still observed but {does not} go exactly
to zero. For {the} cuprates superconductors{,} with doping around
$12\%$, a softening of the phonon dispersion line is clearly observed,
but, surprisingly, {instead of taking place} below $T_{co}$, {it
occurs} below $T_{c}$. It is {very} unusual since there is no
specific reason why the CDW modulations {should} be sensitive to
the {superconducting} ordering temperature $T_{c}${,} rather
than $T_{co}$. In this theoretical scenario, we account for the sensitivity
of the phonon line to $T_{c}$, by considering that the phase freezing
of the CDW and SC modes both occur below the minimum of the two temperatures
$T_{c}$ and $T_{co}$, which here is $T_{c}$. Above $T_{c}${,}
the phases of the particle-particle and particle-hole order parameters
{unlock} and fluctuate, which {wash} out the softening of the
line~\cite{Sarkar21}. }

\subsection{{Cascade of phase transitions, loop currents}}

{One of our early {discoveries} was that a cascade
of phase transitions could be produced around an AF {quantum critical
point} (QCP), {thus} masking the QCP itself ~\cite{Meier14}.
This can be one interpretation of the experimental phase diagram of
{the} cuprates, where spin-glass, charge ordering, {time-reversal}
(TR) symmetry breaking, {inversion-symmetry (IS) breaking}, and
{loop currents} have been reported in the vicinity of the AF QCP.
Out of this complexity, one can already notice a body of phase {transitions}
involving discrete symmetries and not breaking translational symmetries.
Since they {do not} break {Galilean} invariance, they cannot be
directly responsible for the opening of the PG in the {antinodal}
(AN) region, but experiments show that in the {underdoped} region
they ``accompany'' the formation of the PG.}

\begin{figure}
{\includegraphics[width=8cm]{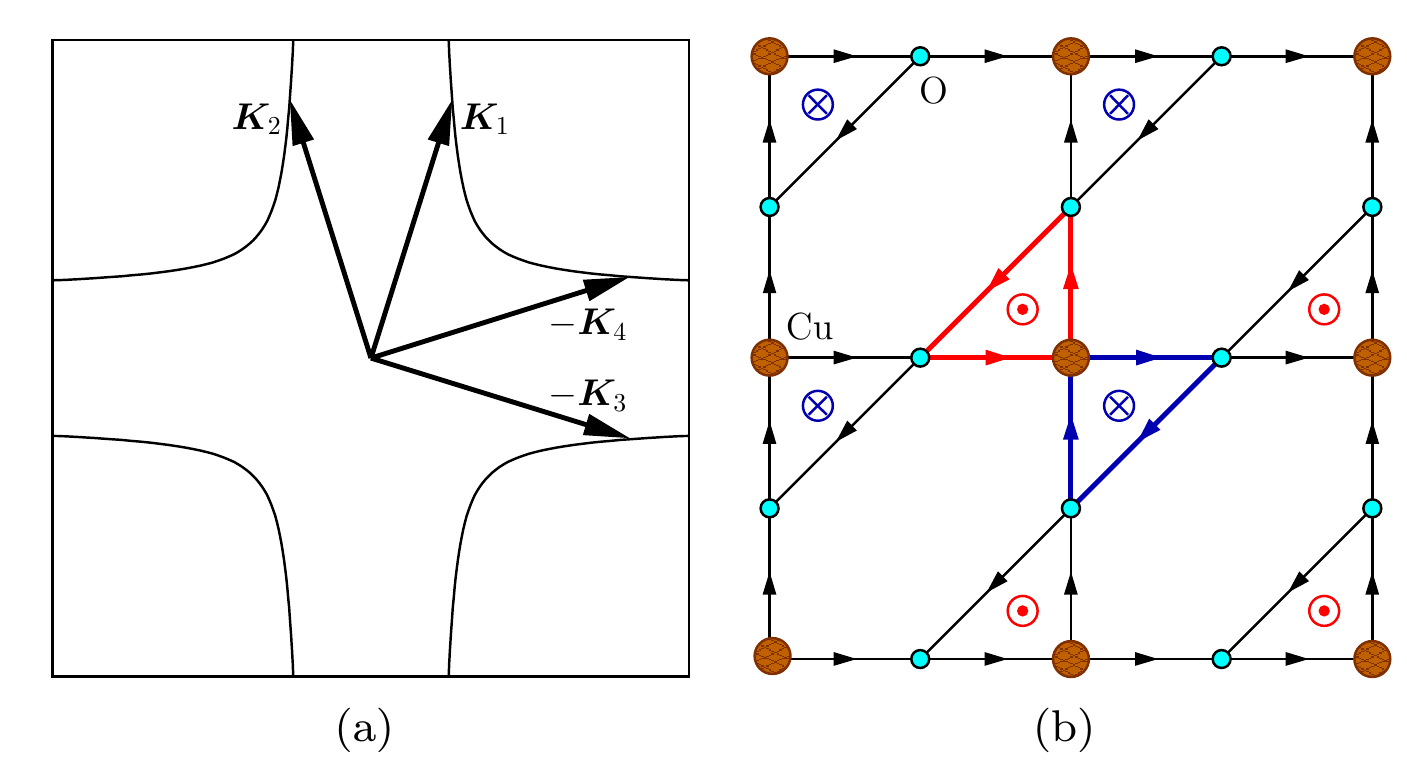} \caption{\label{fig:Schematic-illustration-of}Schematic illustration of the
loop currents. From Ref.~\cite{Agterberg:2014wf}.}
}
\end{figure}

{We focus here on the formation of the {loop currents}
(LC) and see how they can integrate into our theoretical framework~\cite{Carvalho15,Carvalho16,Agterberg:2014wf}.
As depicted in Fig. \ref{fig:Schematic-illustration-of}, loop currents
are a structure observed by elastic neutron scattering, namely{,}
at $\mathbf{Q}=0$ wave vector. They {do not} break translational
invariance and thus are notoriously difficult {to measure}. They
break both TR and IS, but not the product of the two. There are two
approaches to the issue of LC order in {the} cuprates. In the first
approach{,} the LC order comes directly from the three{-}band
Hubbard model as an exotic, but very {important} primary order.
The second way is to consider that the LC order is a ``vestigial''
order, namely{,} a discrete order that {comes} as a precursor
of a continuous phase transition. One then needs to check {if} the
discrete order is compatible with the main continuous order as far
as symmetries are concerned. A notable candidate for the precursor
order is the {pair density wave} (PDW) order, which allows for the
formation of {particle-particle pairs} with finite center{-}of{-}mass
momentum~\cite{Agterberg:2014wf}. In Ref.~\cite{Sarkar19}, we
discuss the idea of LC as a precursor order in the context of a fractionalized
PDW scenario. Our conclusion is that the symmetry of fractionalized
PDW allows for LC as a precursor order.}

{Due to the fact that the LC order is a $\mathbf{Q}=0$
order, the experimental evidence for its presence through elastic
neutron scattering is difficult and has been controversial since a
lot of subtractions have to be made to the signal to finally extract
the response corresponding to the LC~\cite{Fauque06,Bourges18,Bourges11}.
Recent developments, nevertheless{,} reveal the presence of a signal
at finite wave vectors $\left(0,\pm\pi\right)$ and $\left(\pm\pi,0\right)$.
The presence of LC at {a} finite wave vector being a clear signature
of the signal gives hope for a resolution of the experimental controversy
in {the} near future.}

{A question that is often debated with the idea of
LC is that of the nature of the transition at $T^{*}$ . It is important
to note that the LC is not a static ``order''{,} but rather a
dynamical phenomenon. For one thing{,} they are not visible by NMR,
which shall be the case for a static magnetic order since the typical
timescale of NMR is $10^{-6}$ s. For comparison, the timescale of
{neutron scattering} is $10^{-9}$ s, while the one for muon scattering
resonance is ($\mu$SR) $10^{-12}$ s. A signal corresponding to LC
has been reported in $\mu$SR experiments, supporting the dynamical
nature of this phenomenon. The transition at $T^{*}$ even from the
point of view of a discrete order like LC is a dynamic phenomenon
which is closer to a {crossover} {rather} than a phase transition.
}

{Another concept {highly} debated around the LC
order and discrete phase transitions {is} whether the phase diagram
has a {quantum critical point} (QCP) around optimal doping, precisely
at the point where the LC order and other discrete phase transition
line seem to terminate at $T=0$. Here, the experiments are a bit
contradictory with both {ARPES} spectroscopy and Raman {experiments,}
which seem to show an abrupt end of the PG line around optimal doping~\cite{Benhabib:2015ds}.
The number of charge carriers was {studied} by Hall conductivity
by one experimental group in YBCO, Nd-LSCO and LSCO~\cite{Badoux16,Badoux:2016kg}.
They find a transition around optimal doping for a number of carriers
{going} from $p$ to $1+p${,} where $p$ is the oxygen doping,
suggesting a QCP at a critical doping $p^{*}$~\cite{Morice:2017kd,putzke2019reduced}.
Another experimental group studied the same quantity in Tl2201 and
Bi2201 and{,} although they converged in finding that the number
of carriers goes form $p$ to $1+p$, the transition between the two
regimes looks more like a wide region around the critical doping where
another type of charge carrier seems to coexist with electronic carriers
~\cite{putzke2019reduced,Ayres}. We will come back {to} these
different viewpoints in our study of transport in the strange metal
phase of those compounds.}

\section{{Strange Metal \label{sec:Strange-Metal}}}

\subsection{{{Longitudinal conductivity and} Hall conductivity}}

\begin{figure}
{\includegraphics[width=8cm]{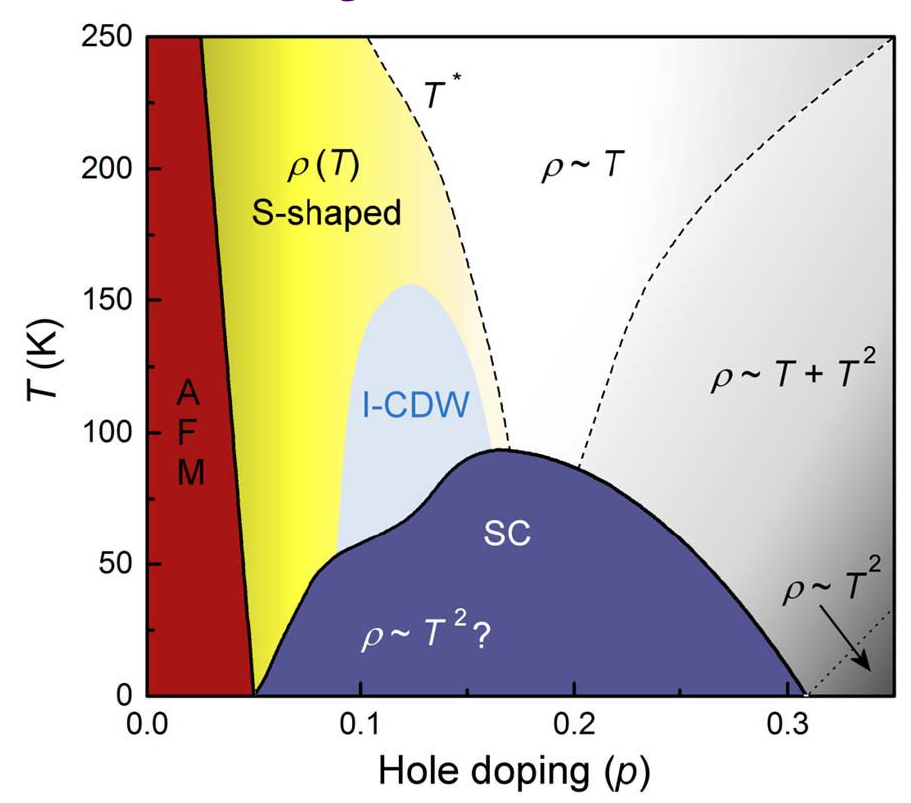} \includegraphics[width=8cm]{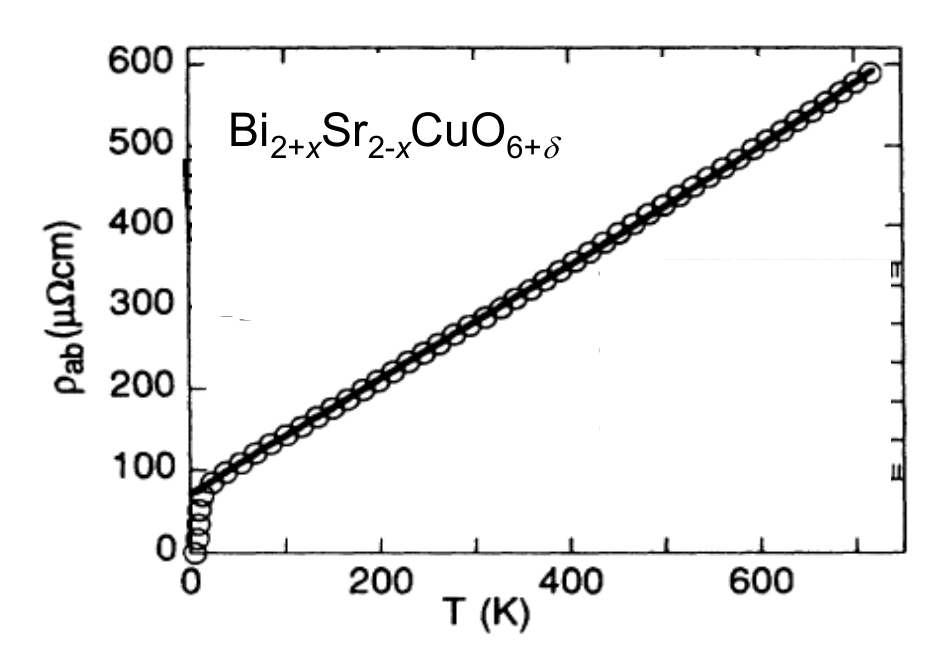}
\includegraphics[width=8cm]{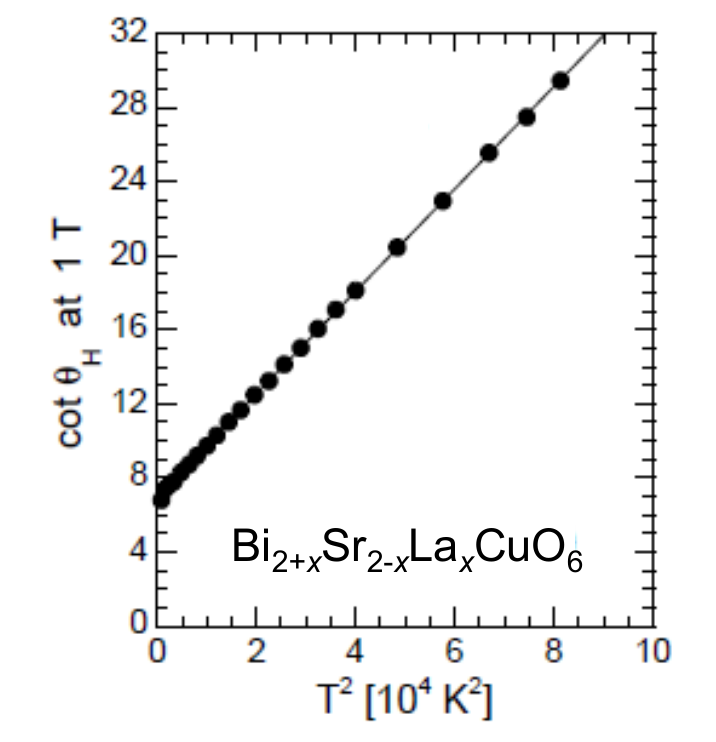} \caption{\label{fig:Schematic-picture-of}Schematic picture of the phase diagram
for transport measurements. The strange metal regime shows the linear
in {$T$} resistivity behavior. {In the lower panels,} the experimental
data for the longitudinal resistivity and the $\cot\left(\theta_{H}\right)$
varying {with} $T^{2}$. {From Ref.~\cite{hussey2008phenomenology}.}}
}
\end{figure}

{The strange metal phase of {the} cuprates is one
of the greatest mysteries of condensed matter physics. Around the
optimal doping, the longitudinal resistivity is linear in temperature
($\rho_{xx}\sim T$) over several decades of temperature{,} while
the Hall cotangent, which is the ratio of the linear resistivity to
the transverse resistivity, varies as the square of the temperature
$\cot\theta_{H}=\sigma_{xx}/\sigma_{xy}\sim T^{2}$ (see{, e.g.,}
\cite{hussey2008phenomenology,hussey2013generic} and {references}
therein, and Fig. \ref{fig:Schematic-picture-of}). In a theory where
only one species of charge carrier is present, these two quantities
should be inversely proportional to the {transport lifetime} of
the charge carrier, and thus the same power law should be observed.
Moreover{,} it has been shown that the optical conductivity has
a Drude form as a function of energy, with the Drude width being of
the order of $k_{B}T$. This set of {observations} is notoriously
difficult to account for theoretically. Recently{,} theories have
emerged which consider the special case of optimal doping to be one
of the most strongly {entangled} fixed points that nature has ever
produced. Models using Sachdev-Ye-Kitaev (SYK) Hamiltonians are studied
to describe such a {strongly-correlated} fixed point~\cite{hartnoll2018holographic}.
}

\subsection{{The scenario of entangled bosons}}

\begin{figure}
{\includegraphics[width=8cm]{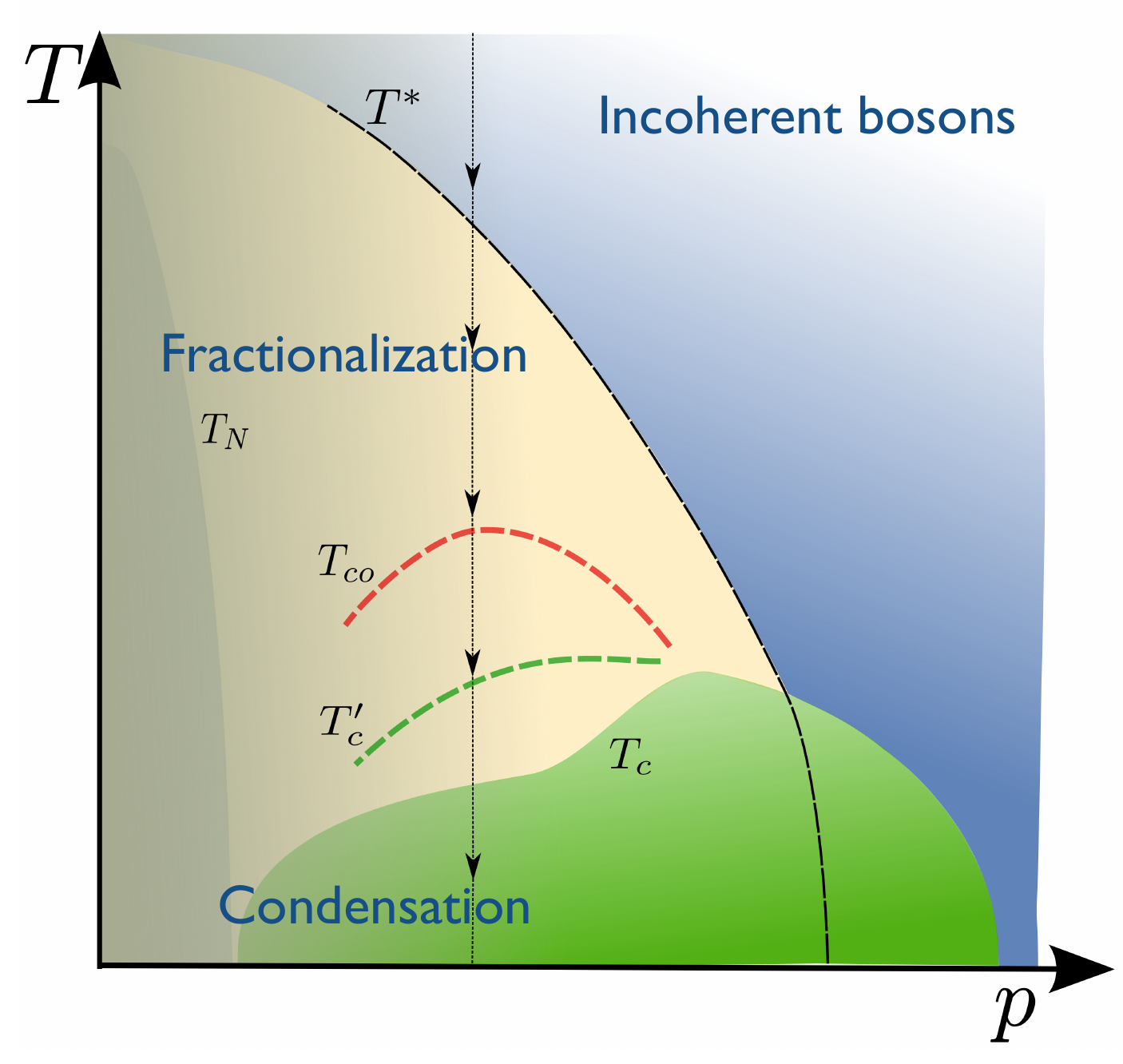} \caption{\label{fig:Phase-diagram-of}Phase diagram of the cuprates within
the fractionalized scenario. At high energy, due to strong coupling,
incoherent bosons are formed. At $T^{*}${,} some bosons like the
PDW {do not} \textquotedblleft survive\textquotedblright{} and fractionalize.
{At lower temperatures,} particle-hole pairs can condense, forming
the charge orders observed {in the underdoped regime} of the phase
diagram. From Ref.~\cite{Banerjee21}.}
}
\end{figure}

{In our scenario, we take a new approach based on
recent transport experiments that suggest the presence of a new carrier
species in a doping region around the optimal doping. This species
would participate to the linear resistivity in temperature but would
not contribute to the Hall conductivity. Moreover{,} a scaling relation
in $H/T${,} where {$H$} is the applied magnetic field has been
observed. The picture that emerges (and is depicted in Fig. \ref{fig:Phase-diagram-of})
for the cuprates is that at ``high energy'' with $0.5$ eV $<\omega<1$
eV, many bosons form due to strong coupling interactions. Many symmetry
channels are represented with spin 1, spin 0, zero charge or charge
two bosons, and they form at different wave vectors. At strong coupling{,}
these bosons have a {``hard core''}, with the pairing energy being
approximately equal to the Coulomb energy. When the temperature decreases,
some bosons are unstable, like the charge-{two} bosons with finite
wave vector (which are remnants of PDW order){. These bosons} ``fractionalize''
and some other bosons condense around the optimal doping, like the
particle-hole bosons of zero charge and finite wave vector, producing
charge density waves. A recent experiment that supports this picture
is the MEELS experiment at optimal doping for {the} cuprates. We
see that the compressibility as a function of frequency saturates
to a plateau at low frequency, whose width is of the order of 0.7
eV. Scalings relations are also shown experimentally. These data suggest
that we are dealing here with a {``jamming''} transition{,}
which appears naturally if we have interacting bosons near the optimal
doping \cite{Pierfrancesco2022}.}

{In two papers~\cite{Banerjee20,Pangburn2022},
we {calculated} the conductivity and thermal conductivity of {the}
boson-fermion ``soup'' with charge-two bosons. In this picture,
the bosons scatter through the fermions and become Landau-damped {described
by} 
\begin{align}
D_{\mathbf{q},\omega}^{-1} & =\left|\omega\right|+\mathbf{q}^{2}+\mu(T),\label{eq:7}
\end{align}
with $\mu\left(T\right)$ {being} the ``bosonic mass''. {Besides,}
$\mu\left(T\right)\sim gT+\mu_{0}${,} where the term proportional
to the temperature is due to the bosons interacting together and $\mu_{0}=0$
when the bosons are critical. Since this theory is particle-hole symmetric,
bosons with propagators described by Eq. (\ref{eq:7}) {do not}
contribute to the transverse conductivity and thus {do not} affect
the Hall conductivity. We are thus oriented by a global picture where
two types of carriers are present. Bosons contribute to the temperature
linear resistivity but not to the Hall conductivity. On the other
hand, fermions will contribute to the Hall conductivity and{,} in
a model with ``hot'' and ``cold'' regions{,} a phenomenological
study has shown that fermions contribute to the Hall conductivity
as $\sigma_{xy}\sim T^{-3}$~\cite{Hussey_McKenzie}. Assuming that
the bosons short-circuit the fermions and contribute to the longitudinal
conductivity as $\sigma_{xx}\sim T^{-1}${,} one gets the correct
scaling for the cotangent Hall $\cot\theta_{H}\sim T^{2}$.}

{We have employed a Kubo formulation of the transport
for the bosons~\cite{Banerjee20} and also a more hydrodynamic formulation
{using the memory-matrix approach} \cite{Pangburn2022}, and found
that indeed {such} bosons contribute to the linear in {$T$} resistivity
in $d=2$. {To obtain this result,} we {assumed Umklapp scattering
as the mechanism for total momentum relaxation in the theory}. }

\section{{Discussion\label{sec:Discussion}}}

{Before concluding this paper, it is worth discussing
the view of other works on the issue of charge order in the underdoped
region of the cuprate superconductors. As mentioned in the section
\ref{subsec:Raman-scattering-and}, some researchers believe that
the charge order and its fluctuations are key for understanding the
PG phase, whereas others think it is an epiphenomenon. Among the works
that propose that charge order is important, we would like to mention
a series of studies where the fluctuations of the charge order are
seen over a wide range of the phase diagram \cite{Ghiringhelli12,GhiringhelliArpaia,Ghiringhelli19}
and scattering through charge modes are thought to explain the linear-in-$T$
resistivity in the strange metal phase \cite{Caprara:2016gs,Caprara:2016vh}.
This way of thinking is closer to our approach, although the concept
of emergent symmetry is not present. Then, a group of other researchers
view the PG state as a fluctuating PDW \cite{agterberg2020physics}.
Two approaches are given: In one of them, the PDW is the ``mother''
state from which the charge order emerges at lower doping \cite{Lee14}.
In the other approach, the charge order is a vestigial state coming
out of the PDW fluctuations \cite{Fradkin15}. Both approaches have
in common with ours that the PDW fluctuations are important in the
phase diagram of the cuprates. However, they differ in that the PDW
order parameter fractionalizes in our approach. Most of the other
approaches to the charge order in the cuprates consider that it is
a small phenomenon, which cannot be important to understand the PG
physics (see, e.g., \cite{Fradkin15} and references therein). Nevertheless studies of CDW in under-doped cuprates although de-correlated from the PG, have led to interesting discussion about its origin and the fact that the charge ordering wave vector is not situated on the diagonal fin the Brillouin zone (see, e.g., Ref.~\cite{Wang14}). In
particular, non-magnetic charge order has not yet been systematically
found numerically in the underdoped region (see, e.g.,~\cite{Corboz14}).
It is still an open question as to whether the numerical accuracy
will improve enough in the near future to finally settle the issue
of the importance of charge order in underdoped cuprates.}

{In this paper{,} we have given an overview of
the PG and the strange metal phase of {the} cuprate superconductors
with a special focus on the charge order in the underdoped region.
The main idea is that the charge order is in competition with superconductivity,
but also, in a surprising and unprecedented way in condensed matter,
the two orders {turn out to} be entangled. These mixed effects of
entanglement and competition are described by chiral models, in which
the square of the amplitudes of the two modes are linked by a local
constraint. The first chiral model introduced to describe the PG is
the {non-linear O(3) $\sigma$} model. In this context{,} it corresponds
to the idea of an emergent SU(2) symmetry{,} where $\eta$-modes
describe a rotation from the superconducting state to the charge order
state. An exact realization of the SU(2) symmetry was found and treated
in the {eight-hot-spot} model. The constraint linking the two orders
form a {pseudospin} space. This type of model allows to explain
the opening of the PG in the {antinodal} part of the Brillouin zone
and the formation of Fermi arcs. We obtain a good description of the
PG {opening} observed by ARPES in Bi2201, which takes into account
the back-bending of the electronic dispersion around a wave vector
$\mathbf{Q}_{0}/2>\mathbf{k}_{F}$, where $\mathbf{k}_{F}$ is the
Fermi wave vector. The phase diagram as a function of magnetic field
and temperature is also very well described. In particular{,} we
hold that the flat transition occurring at {$B=18$ T} due to a
three-dimensional charge order cannot be described by a Ginzburg-Landau
model, but requires a constraint. This transition is {well described}
by the {non-linear $\sigma$} model{,} which captures the flatness
of the transition, interpreted here as a pseudo-spin-flop transition
by analogy with the ``spin-flop'' transitions observed in spin models.
A recent Raman diffusion experiment also goes in the same direction
with the observation of the ``charge gap'' in the $B_{2g}$ channel.
Strikingly{,} the charge gap is of the same order of magnitude as
the observed SC gap in the $B_{1g}$ channel. Moreover, it follows
$T^{*}$ with the oxygen doping, rather than following the charge
ordering temperature $T_{co}$. In this type of interpretation, the
{occurrence} of charge modulations inside the vortex core upon applying
an {external} magnetic field is a topological object: a skyrmion
in the {pseudospin} space. The presence of $\eta$-modes in the
system can also be related to an x-ray experiment showing some optical
modes around $\mathbf{Q}_{0}$ interacting with a phonon.}

{In {the} second part, we {introduced} the concept
of fractionalization of an order associated with a density wave of
Cooper pairs. The idea of fractionalization is not new, and was introduced
at the beginning of the study of {the} cuprate superconductors with
the original paper of P. {W.} Anderson's resonating valence bond~\cite{Anderson87}.
This idea was implemented in large part in gauge theories for which
the electron is fractionalized into its spinon and holon elementary
components (see, e.g.,~\cite{Lee06}). Here, we take up the same idea
of fractionalization, but on a collective order-parameter field: a
Cooper pair density wave (PDW). The PDW fractionalizes at {$T^{*}$}
into a particle-particle field
and a modulated particle-hole field, breaking the translational invariance.
As before, the PG state is described by these two fields, which are
both intertwined and in competition. The cuprate phase diagram is
now interpreted {as} a deconfinement transition at $T^{*}$ with
decreasing temperature and the quantum rotor model describes the physics
under $T^{*}$. This approach has given us a way to think about the
strange metal phase observed at optimal doping in {the} cuprates.
The idea is that at a high energy scale, of the order of 0.7 eV and
under the effect of the very strong {Coulomb} interaction, bosons
(associated with the order-parameter field) emerge with different
symmetries. Some have charge two, some {have} charge zero, others
are spin singlet{,} etc. These bosons, coupled to the lattice, can hop from site to site and thus have a dispersion. When the temperature
{is lowered}, some bosons will be unstable and will fractionalize,
while others will be stable and will condense, giving rise to charge
orders at different {locations} in the phase diagram. {If we consider}
that bosons of charge-two are present in the strange metal phase,
we have shown that they {will} contribute with a {$T$-}linear
resistivity, thus shedding {potentially} new light on this {big}
mystery of quantum material physics.}

\section{{Acknowledgments}}

{This paper is dedicated to late Konstantin B. Efetov
with gratitude for the very creative research undertaken together
and with nostalgia for the wonderful time spent working in various
places of the world. Kostya was a rare physicist, independent, inventive
and brilliant. The community is missing him sorely.}

{We would like to thank the many young physicists
who participated in this project at various stages of its development:
Anurag Banerjee, Corentin Morice, Debmalya Chakraborty, Emile Pangburn,
Ervand Kandelaki, Hendrik Meier, Louis Haurie, Matthias Einenken,
Maxence Grandadam, Saheli Sarkar, Thomas Kloss, Vanuildo S. de Carvalho
and Xavier Montiel. We acknowledge the International Institute of
Physics (IIP) in Natal (Brazil), where part of the collaboration with
Kostya was happening. A special thanks to A. Ferraz, its director,
for his kind hospitality. Discussions with experimentalists were invaluable
for the ideas exposed in this work. We are especially grateful to
Y. Sidis for numerous and often heated discussions about the experiments.
We also acknowledge very useful interactions with H. Alloul, V. Balédent,
D. Bounoua, P. Bourges, D. Colson, S. C. Davis, Y. Gallais, G. Grissonnanche,
M. Hamidian, N. Hussey, M. -H. Julien, D. Leboeuf, B. Leridon, M.
Le Tacon, M. -A. Méasson, C. Proust, B. Ramshaw, A. Sacuto, S. E.
Sebastian, L. Taillefer and P. Urbani. H.F. also acknowledges funding
from CNPq under Grant No. 311428/2021-5.}

\bibliographystyle{pnas2009}
\bibliography{Cuprates}

\end{document}